\documentclass[aps,prb,groupedaddress,nolongbibliography,twocolumn]{revtex4-2}
\usepackage{amsmath,amssymb}
\usepackage{graphicx,xcolor,soul,sidecap}
\usepackage[colorlinks,bookmarks=false,citecolor=blue,linkcolor=blue,urlcolor=blue,hyperfootnotes=true]{hyperref}
\usepackage{dcolumn}
\usepackage{bm}

\begin{document}

\preprint{APS/123-QED}

\title{Evidence for a competition between the superconducting spin-valve effect and quasiparticle spin-decay in superconducting spin-valves}

\author{B. Stoddart-Stones}
\email{bs507@cam.ac.uk}
\author{X. Montiel}
\author{M. G. Blamire}
\author{J. W. A. Robinson}
\email{jjr33@cam.ac.uk}
\affiliation{%
 Department of Materials Science \& Metallurgy, University of Cambridge, 27 Charles Babbage Road, Cambridge CB3 0FS, United Kingdom
}%

\date{\today}

\begin{abstract}
{The difference in the density of states for up- and down-spin electrons in a ferromagnet (F) results in spin-dependent scattering of electrons at a ferromagnet / nonmagnetic (F/N) interface. In a F/N/F spin-valve, this causes a current-independent difference in resistance ($\Delta R$) between antiparallel (AP) and parallel (P) magnetization states. Giant magnetoresistance (GMR), $\Delta R = R(AP) - R(P)$, is positive due to increased scattering of majority and minority spin-electrons in the AP-state. If N is substituted for a superconductor (S), there exists a competition between GMR and the superconducting spin-valve effect: in the AP-state the net magnetic exchange field acting on S is lowered and the superconductivity is reinforced meaning $R(AP)$ decreases. For current-perpendicular-to-plane (CPP) spin-valves, existing experimental studies show that GMR dominates ($\Delta R>0$) over the superconducting spin valve effect ($\Delta R<0$)  [J. Y. Gu \textit{et al.}, Phys. Rev. B 66, 140507(R) (2002)]. Here, however, we report a crossover from GMR ($\Delta R > 0$) to the superconducting spin valve effect ($\Delta R < 0$) in CPP F/S/F spin-valves as the superconductor thickness decreases below a critical value.
}
\end{abstract}

\maketitle
\section{Introduction}
The field of spintronics \cite{Zutic2004} emerged following the discovery of spin-dependent scattering of electrons at ferromagnetic/nonmagnetic (F/N) interfaces \cite{Johnson1985} and giant magnetoresistance (GMR) in F/N/F structures \cite{Dieny1991}. In a F/N/F spin-valve, GMR is the difference in electrical resistance ($\Delta R$) between antiparallel (AP) and parallel (P) magnetisation states of the F layers and is current-bias independent. In the AP-state, both the majority and minority spin-electrons are strongly scattered and $\Delta R = R(AP)-R(P)>0$ with the magnitude of $\Delta R$ dependent on the spin-polarization of the F layers, interfacial spin-flip, and the spin decay length in N  \cite{Valet1993,Bass2016}. 

In superconducting F/S/F spin-valves \cite{Oh1997,Tagirov1999,Buzdin1999}, the superconducting critical temperature ($T_\text{c}$) depends on the magnetic moment orientation of the F layers due to the superconducting spin valve effect: in the P-state, the magnetic exchange fields suppress $T_\text{c}$(P) relative to $T_\text{c}$(AP) in which the magnetic exchange fields partially cancel meaning $\Delta T_\text{c}=T_\text{c}^{AP}-T_\text{c}^{P} > 0$. This effect allows the superconducting spin-valve to act as a valve for superconducting current flow, demonstrating infinite magnetoresistance, via switching the magnetic state of a device with suitably large  $\Delta T_\text{c}$ held at constant temperature. This current-bias independent behaviour is observed in current-in-plane (CIP) F/S/F spin-valves with $\Delta T_\text{c}$ reaching tens of mK for transition metal Fs \cite{Gu2002,Potenza2005,Moraru2006,Moraru2006a,Miao2008,Zhu2009,Leksin2010,Zhu2010,Leksin2015,Jara2019} and several hundred mK for rare-earth ferromagnetic metals and insulators \cite{Gu2015,Li2013,Zhu2016}. These experimental values of $\Delta T_\text{c}$ are orders of magnitude smaller than values predicted by theory \cite{Oh1997,Tagirov1999,Buzdin1999}, as it has proven experimentally challenging to reach the theoretically indicated optimum parameter space. Negative $\Delta T_\text{c}$ values have also been reported \cite{Zhu2009,Aarts2006,Rusanov2006,Steiner2006,Singh2007,Singh2007a,Stamopoulos2007a,Leksin2009,Hwang2010,Flokstra2010a}, attributed either to quasiparticle (QP) spin-accumulation \cite{Rusanov2006,Singh2007a,Leksin2009} suppressing $T_\text{c}$ in the AP-state \cite{Takahashi1999,Takahashi2003}, or flux penetration in S from out-of-plane domain walls in the F layers  \cite{Zhu2009,Steiner2006,Stamopoulos2007a,Flokstra2010a}.
 
\begin{figure}
    \includegraphics[width=0.9\linewidth]{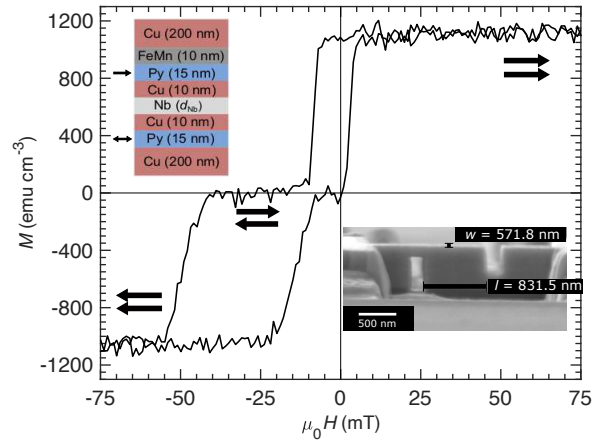} 
    \caption{\label{fig:pillar_withMH} Magnetization ($M$) vs. in-plane magnetic field ($H$) hysteresis loop for an unpatterned spin-valve at 10~K (arrows indicate the net magnetic moment directions of the top and bottom Py layers). Inset top left: Schematic diagram of the superconducting spin-valve. Arrows represent pinned (top) and free (bottom) ferromagnetic layers. Inset bottom right: Scanning electron micrograph of a nanopillar spin-valve. `Length' ($l = 831.5\pm40$~nm) and `width' ($w = 571.8\pm40$~nm) of the device are labelled. Device area, $A = l\times w = 4.7\pm0.4 \times 10^5$~nm\textsuperscript{2}. }
\end{figure}

Current-perpendicular-to-plane (CPP) devices have larger values of GMR than CIP devices \cite{Pratt1991,Lee1995,Gijs1995,Eid2002}, but are less investigated due to the extra complications of fabrication compared to CIP spin-valves, and so investigation into CPP devices with superconducting spin-valves has been limited. One reported CPP device \cite{Gu2002a} was a F/S/F spin-valve, which showed GMR behaviour ($\Delta R >0$) due to quasiparticle transport (``QP GMR") with a reduced spin decay length relative to the normal state for superconducting Nb layer thicknesses exceeding 30~nm. We note that the superconducting spin-valve effect ($\Delta R <0$) was not observed and that superconducting devices with thicknesses below 30~nm were not reported in \cite{Gu2002a}. \par
In this article, we systematically investigate superconducting CPP F/S/F spin-valves with Py(15)/Cu(10)/Nb($d_\text{Nb}$)/Cu(10)/Py(15)/FeMn(10) layers (numbers in nm units) sandwiched between 200-nm-thick Cu electrodes (Fig.~\ref{fig:pillar_withMH}, inset top left). As expected, with decreasing Nb thickness ($d_\text{Nb}$) QP GMR increases; however, below a critical thickness of superconducting Nb ($d_\text{Nb} = 26$~nm) a sign change in $\Delta R$ is observed, consistent with the appearance of the superconducting spin valve effect in these CPP devices, which dominates the QP GMR behaviour at these thicknesses. We show a systematic crossover between these competing behaviours, dependent on $d_\text{Nb}$. 

\section{Experimental}
The spin-valves are prepared by dc magnetron sputtering in an ultra-high vacuum chamber with a base pressure better than $10^{-8}$~mbar. Films are deposited onto single crystal silicon with a 250-nm-thick surface oxide, with an in-plane magnetic field (100~mT) applied during growth to set in-plane uniaxial anisotropy. An antiferromagnetic layer of FeMn exchange biases the top layer of Py (Ni\textsubscript{80}Fe\textsubscript{20}), ensuring a stable AP-state (Fig.~\ref{fig:pillar_withMH}). The Cu between the Nb and Py improves interface quality by limiting magnetic dead layers \cite{Bell2005,Robinson2007,Tateishi2010}, and also increases the magnitude of magnetoresistance \cite{Gu2002a}. The spin-valves are patterned into nanopillars using optical lithography and Ga-ion focused ion-beam (FIB) etching (described elsewhere~\cite{Bell2003a}). The lengths and widths of the nanopillars vary between 400-1500~nm and 300-1000~nm, respectively. Dimension and resistance values for each device are reported in Table~S1 \cite{SM}.
Resistance of the CPP spin-valves vs. temperature [$R(T)$] or in-plane magnetic field [$R(H)$] is measured using a `quasi' four-point current-bias setup in a pulse-tube measurement system, which removes contact resistance. As Cu contact layers are used, the section under measurement is not exclusively the nanopillar device, but includes part of the patterned structure from which the nanopillar was milled, which is $4~\mu$m wide and $20~\mu$m long. This section of heterostructure, referred to as the `contact leads', contributes to the measured resistance and so we refer to the measurement as `quasi' four-point. Finally, in CPP measurements, since the cross-sectional areas ($A$) of the CPP spin-valves vary, $\Delta R$ is normalized by multiplying by $A$ (i.e. $A\Delta R$) \cite{Bass2016}. Measurements were made on devices in both the normal state (above the superconducting transition of the nanopillar device and contact leads, but  below $10$~K) and in the superconducting state, although many devices had such a suppressed transition temperature that they could not be measured in the S state. One important impact of the contact lead resistance is that $R(T)$ measurements can appear to contain two distinct superconducting transitions [Fig.~\ref{fig:RTandRH}(a,d)]. The higher temperature transition (`contact transition') corresponds to the Nb in the contact leads, whereas the lower temperature transition (`device transition') corresponds to the nanopillar device. We define the onset temperature of the latter as $T_\text{device}$, and $T_\text{c}$ as the temperature at 50\% of the resistance change below $T_\text{device}$ (Supplementary Material Section S-IIB \cite{SM}).

\section{Results}
\begin{figure*}
    \includegraphics[width=0.9\linewidth]{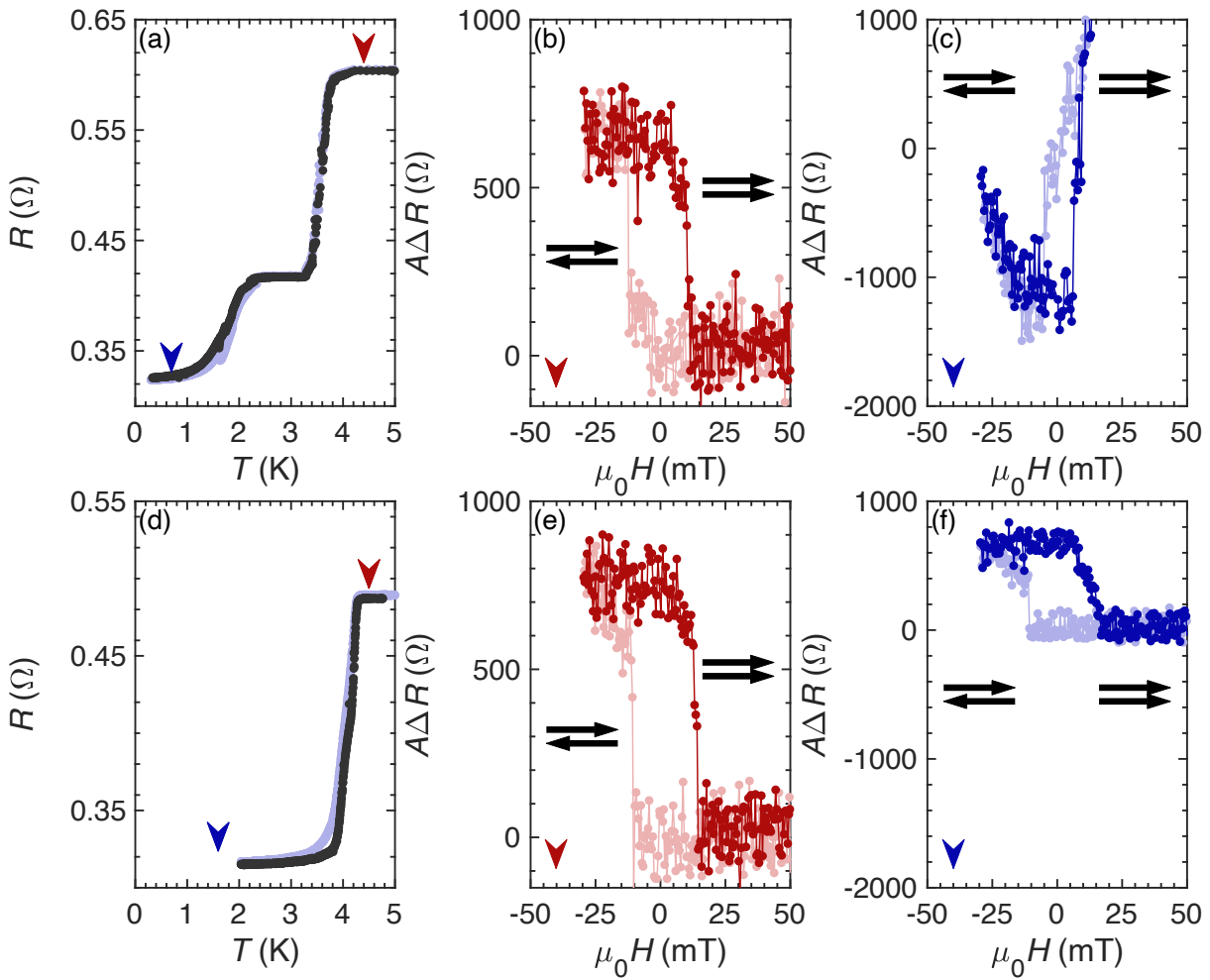}
    \caption{The two different device responses observed. (a,d) $R(T)$ curves for two different devices  (top, $d_\text{Nb} = 25$~nm; bottom, $d_\text{Nb} = 28.5$~nm) in both the parallel (black) and antiparallel (blue) states. (b,e) Normal state minor $R(H)$ loops from the same devices, both demonstrating GMR. (c,f) Minor $R(H)$ loops in the superconducting state. The device with $d_\text{Nb} = 25$~nm has greater resistance in the parallel state when superconducting, whereas the device with $d_\text{Nb} = 28.5$~nm has greater resistance in the antiparallel state, similar to the normal state. For the $R(H)$ loops, light data represent sweeps from positive to negative $H$, starting at high positive values.  The dark data in each loop is for the return sweep. In these minor loops, only the free Py layer undergoes switching, illustrated by the black arrows, which show the relative magnetic moment orientation of the Py layers. Arrows in (a,d) indicate the temperature of the corresponding $R(H)$ loops.}
    \label{fig:RTandRH}
\end{figure*}

$R(T)$ and $R(H)$ measurements of two devices are shown in Fig.~\ref{fig:RTandRH}. Whilst the normal state $R(H)$ loops [Fig.~\ref{fig:RTandRH}(b,e)] indicate GMR behaviour as expected, the superconducting state loops  [Fig.~\ref{fig:RTandRH}(c,f)] reveal that two distinct behaviours appear in our devices, which is supported by the device transitions in Fig.~\ref{fig:RTandRH}(a,d); for the device with higher $d_\text{Nb} = 28.5$~nm, $T_\text{c}(AP)<T_\text{c}(P)$ and $\Delta R>0$, consistent with the results in \cite{Gu2002a}, suggesting GMR moderated by quasiparticles (QP GMR). For the device with lower $d_\text{Nb} = 25$~nm $\Delta R$ is negative [Fig.~\ref{fig:RTandRH}(c)], and $T_\text{c}(P)<T_\text{c}(AP)$. These results are those expected for superconducting spin-valve effect behaviour observed in CIP devices. The resistance increase with increasing field visible in Fig.~\ref{fig:RTandRH}(c) occurs for all superconducting spin-valve effect devices, which all have low $H_\text{c2}$ values (being mid-transition at measurement temperatures), and therefore exhibit visible resistance increases within an applied magnetic field.

\begin{figure}
    \includegraphics[width=0.9\linewidth]{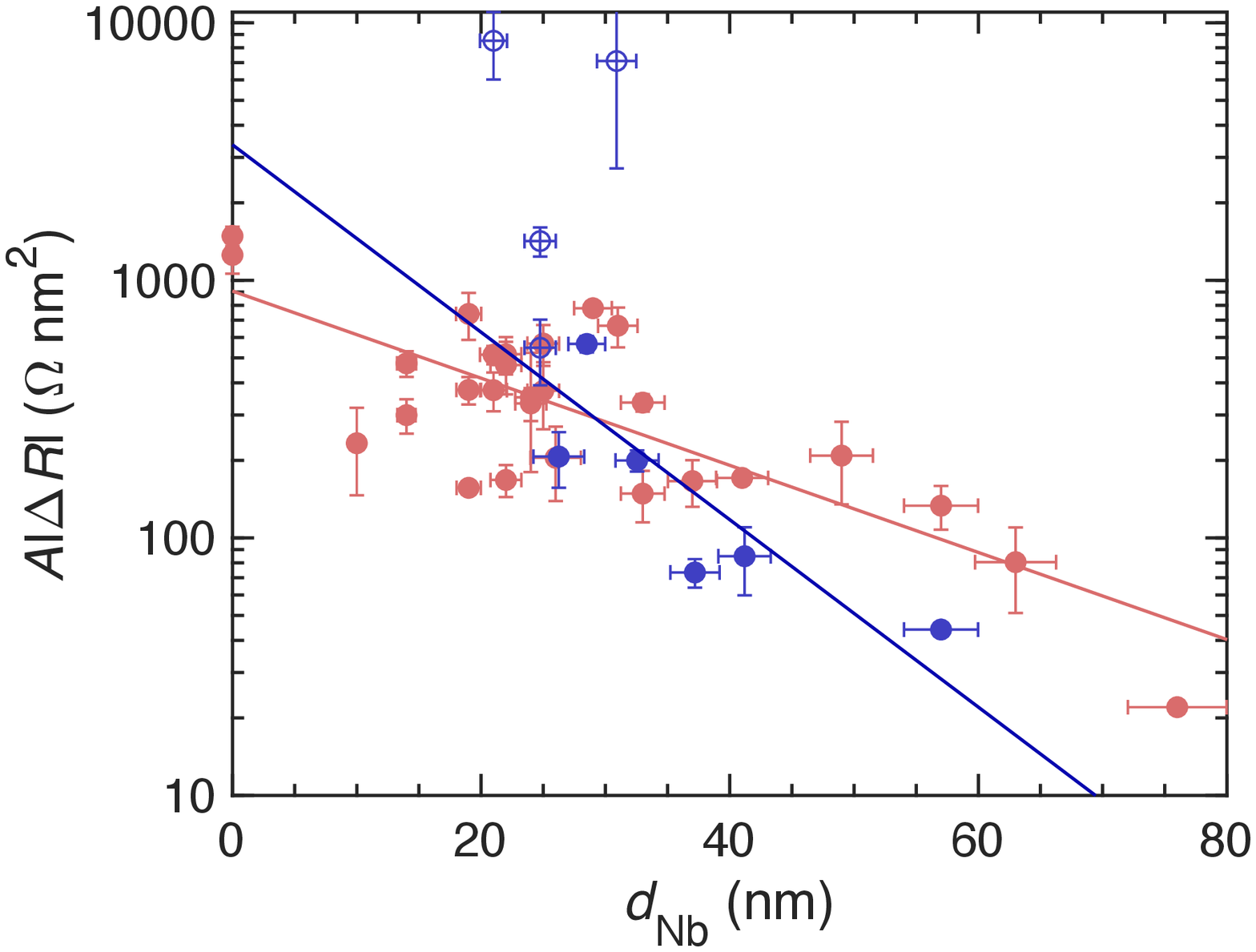}
    \caption{\label{fig:spindecay} $A|\Delta R|$ vs. $d_\text{Nb}$ in CPP spin valves for both the normal state (below 10~K, red) and superconducting state (blue). Closed blue circles are devices showing quasiparticle giant magnetoresistance, whereas open circles show the superconducting spin valve effect. Fits are to a simple exponential decay as described in the text, and open circle points are not included in the superconducting fit. Points are the mean value from multiple devices on a single substrate; vertical error bars are the standard deviation in $A\Delta R$ of these devices and horizontal error bars the uncertainty in $d_\text{Nb}$.}
\end{figure}

In Fig.~\ref{fig:spindecay} we show the absolute value $A |\Delta| R$ \textit{versus} $d_\text{Nb}$, with data points and y-error representing the mean of multiple devices from a single substrate and their standard deviation respectively. Red data are from devices in their normal state (measured at $<10$~K) and blue data are from devices in the superconducting state at $T/T_\text{device} = 0.3$. Open circle data are devices demonstrating the superconducting spin valve effect. By fitting a simple decaying exponential, $\exp{(-d_\text{Nb}/l_\text{sf})}$ \cite{Park2000}, a reasonable approximation where the thickness of the ferromagnet (15~nm) is much greater than the spin-flip length in the ferromagnet (5.5~nm, \cite{Vila2000}), we estimate spin-diffusion lengths from the GMR data in both the normal and superconducting states: $l_\text{sf}^\text{N} = 25\pm3$~nm, and $l_\text{sf}^\text{S} = 12\pm4$~nm. The decay length for QP GMR (in the superconducting state) is shorter than the normal state decay length, as found previously \cite{Gu2002a}.

\subsection{Crossover}
We now detail the main results of this article. To the best of our knowledge, the superconducting spin-valve effect has not been previously observed in CPP superconducting spin valves, and so we investigate the factors that determine the appearance of this effect. 

\begin{figure}
    \includegraphics[width=0.9\linewidth]{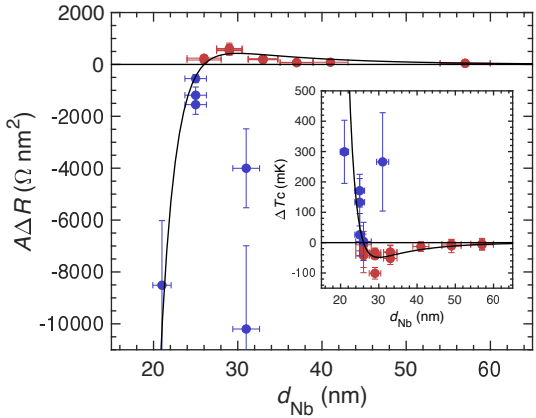}
    \caption{\label{fig:dNb} $A\Delta R$ at $T/T_\text{device}= 0.3$ vs. $d_{\text{Nb}}$, with the inset $\Delta T_\text{c}$ vs. $d_{\text{Nb}}$. Red points show QP GMR dominated behaviour, blue the superconducting spin valve effect dominated behavior. Points represent individual spin-valve devices. Black curves are fits as described in the main text. The crossover point is $d_{\text{Nb}} = 26$~nm.}
\end{figure} 

\begin{figure}
\includegraphics[width=0.90\linewidth]{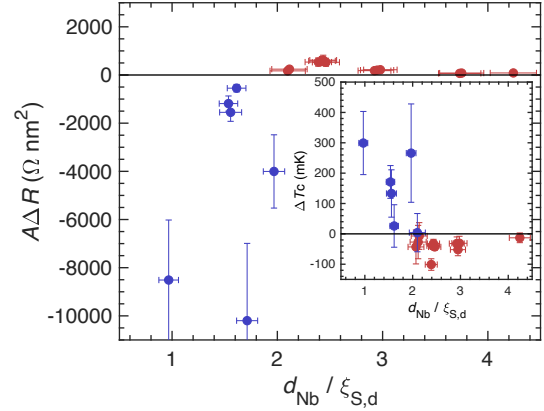}
\caption{\label{fig:dnbbyxi} $A\Delta R$ at $T/T_\text{device}= 0.3$ vs. $d_{\text{Nb}}/\xi_{S,d}$, with inset $\Delta T_\text{c}$ vs. $d_{\text{Nb}}/\xi_{S,d}$. Red points show QP GMR dominated behaviour, blue the superconducting spin valve effect dominated behavior. Points represent individual spin-valve devices. The crossover point between the two behaviours occurs around $d_\text{Nb} = 2\xi_{S,d}$.}
\end{figure}

In Fig.~\ref{fig:dNb} we plot $A\Delta R$ at $T/T_\text{device}= 0.3$ \textit{versus} $d_\text{Nb}$, which shows a systematic dependence of $A\Delta R$ on $d_\text{Nb}$, with a crossover from positive to negative $A\Delta R$ occurring at $d_\text{Nb}=26$~nm. The inset shows the equivalent trend of $\Delta T_\text{c}$ \textit{versus} $d_\text{Nb}$. For $d_\text{Nb}= 21$~nm, the superconducting spin valve effect reaches a positive $\Delta T_\text{c}$ up to 299~mK [Fig.~S1(a) \cite{SM}], which is unprecedented in transition metal F/S/F spin-valves where values are usually of the order tens of mK \cite{Gu2002,Potenza2005,Moraru2006,Moraru2006a,Miao2008,Zhu2009,Leksin2010,Zhu2010,Leksin2015,Jara2019,Aarts2006,Rusanov2006,Steiner2006,Singh2007,Singh2007a,Stamopoulos2007a,Leksin2009,Hwang2010,Flokstra2010a}. However, we note that these large values of $\Delta T_\text{c}$ are linked to inflation of $A\Delta R$; devices showing the superconducting spin-valve effect have suppressed and broadened device transitions [transition width $>1$~K in Fig.~\ref{fig:RTandRH}(a)], meaning even these large $\Delta T_\text{c}$ values will not allow infinite magnetoresistance (complete switching between superconducting or normal states at a constant temperature). 
By considering the impact of superconductivity rather than just $d_\text{Nb}$, using $T_\text{device}$ as the independent parameter, the outlier points at $d_\text{Nb} = 31$~nm were shown to agree far better with the overall trend in the data [Fig.~S2(a) \cite{SM}]. This led to consideration of the value $d_\text{Nb}/\xi_{S,d}$, where $\xi_{S,d}$ is the dirty limit coherence length in the superconductor, calculated using
\begin{equation*}
   \xi_{S,d} = \sqrt{\frac{\hbar D}{1.764 k_\text{B} T_\text{device}}},
\end{equation*}
where $D = 1.4\times 10^{-4}$\,m\textsuperscript{2}s\textsuperscript{-1} is the electron diffusivity, calculated from a coherence length measurement of an isolated 30~nm Nb film, and $k_\text{B}$ is Boltzmann's constant. Using this normalised $d_{\text{Nb}}$, we account for both thickness and processing effects which may affect the superconductivity in the devices. Figure~\ref{fig:dnbbyxi} shows $A\Delta R$ at $T/T_\text{device}= 0.3$ and (Inset) $\Delta T_\text{c}$ vs. $d_\text{Nb}/\xi_{S,d}$, which shows that the crossover between the two behaviours occurs at around $d_\text{Nb} = 2\xi_{S,d}$.\par
Both Fig.~\ref{fig:dNb} and Fig.~\ref{fig:dnbbyxi} show the same overall trend: with decreasing $d_\text{Nb}$, QP GMR  (red) increases as expected, but then peaks and decreases rapidly, devices crossing over into superconducting spin-valve effect dominated behaviour, which rapidly increases in magnitude with decreasing $d_\text{Nb}$. The `peak then fall' shape of the trend indicates that rather than being a sudden switch from QP GMR to superconducting spin-valve effect behaviour, these are two separate effects which compete within the devices. \par

The crossover between positive and negative values of $A\Delta R$ is clear and the magnitude is significant, ruling out minor background effects. Scatter of the data does not account for the crossover behaviour, as indicated by a plot of $\Delta R$ at $T/T_\text{device}= 0.3$ normalised by $\Delta R$ in the normal state [Fig.~S2(b) \cite{SM}]. 

\section{Discussion}
The `double' transition visible within these devices highlights the contribution of the `contact leads' to the measured resistance in these devices. It also highlights that the lower transition - the device transition - tends to be more suppressed in devices demonstrating the superconducting spin-valve effect compared to those demonstrating QP GMR [compare Fig.~\ref{fig:RTandRH}(a) and (d), which feature device transitions around 2~K apart, whereas the `contact' transitions differ by less than 1~K]. 
This observation suggests not only $d_\text{Nb}$, but also the strength of superconducting order within the devices affect the appearance of the superconducting spin-valve effect. \par
\begin{figure}
\includegraphics[width=0.90\linewidth]{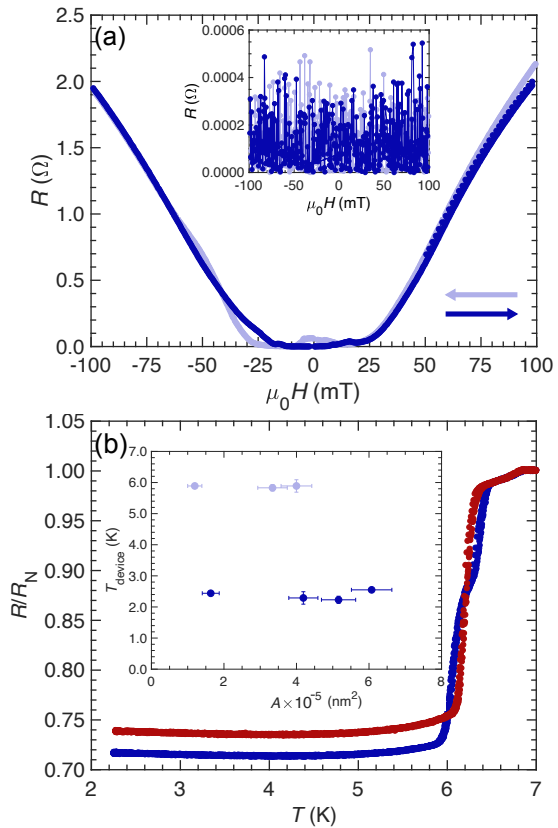}
\caption{\label{fig:wires} (a) Major $R(H)$ loop measured on the contact leads at 0.3~K for the device with $d_\text{Nb} = 21$~nm. For this device only, there is a magnetoresistance response in the contact leads at the measurement temperature, which is negative (parallel state is higher resistance). In all other devices, the leads are superconducting below the contact lead transition, and show no magnetoresistance response at $T/T_\text{device} = 0.3$, as shown in the Inset.  (b) $R(T)$ curves from two devices on the same substrate, normalised by their normal state resistance for comparison. The separate Nb in the nanopillar devices and the contact leads is the origin of the `double transition' visible in $R(T)$ curves. The difference between the temperature of each transition varies in an unknown manner, and can be small enough that the double transition is not distinguishable. Inset: $T_\text{device}$ vs. nanopillar area, for a device showing the superconducting spin valve effect ($d_\text{Nb} = 25$~nm, dark blue) and a device showing QP GMR ($d_\text{Nb} = 37$~nm, light blue). There is no dependence of $T_\text{device}$ on area, which may have been expected due to nanopatterning.}
\end{figure}

Below their superconducting transition, the contact leads are superconducting and do not contribute to the measured $R(H)$ response, as shown in the inset to Fig.~\ref{fig:wires}(a), which was measured on the contact leads only. The exception to this is the substrate with thinnest $d_\text{Nb}=21$~nm: even at the lowest temperatures, the contact leads demonstrate an $R(H)$ response which also shows the superconducting spin valve effect [Fig.~\ref{fig:wires}(a)], which may contribute to the large magnitude of the measured effect for that device.\par
We have considered alternative explanations for the negative $\Delta R$ in our devices [Fig~\ref{fig:RTandRH}(e)] including anisotropic magnetoresistance \cite{Mcguire1975}. However, this is ruled out since anisotropic magnetoresistance is not observed in these superconducting spin-valve effect devices above $T_\text{c}$ [Fig.~\ref{fig:RTandRH}(b)]. Additionally, the zero field $R(T)$ measurements show distinct differences between P- and AP-states and anisotropic magnetoresistance would not lead to such differences in the absence of an applied magnetic field \cite{Dieny1991}. Negative magnetoresistance could also result from crossed Andreev reflection (CAR) \cite{Deutscher2000,Beckmann2004,Beckmann2007,Kleine2009,Webb2012} of electrons across the superconducting layer, when the layer is less than one superconducting coherence length thick. CAR has previously been considered as a source of magnetoresistance in CIP spin valves \cite{Giazotto2006}, but is generally considered as a non-local effect. We note that there is one report of CAR measured in a local setup \cite{Cadden-Zimansky2007}, but the origin of magnetoresistance in this case is uncertain. \par

The values for spin decay length calculated in our devices ($l_\text{sf}^\text{N} = 25\pm3$~nm, and $l_\text{sf}^\text{S} = 12\pm4$~nm) can be compared with other values from the literature. A similar CPP structure has been used previously \cite{Gu2002a} to measure values of $l_\text{sf}^\text{N} = 48\pm3$~nm and $l_\text{sf}^\text{S} = 17.5\pm0.6$~nm. Additionally, a value of $l_\text{sf}^\text{N} = 40\pm5$~nm was extracted from a fit to ferromagnetic resonance data \cite{Jeon2018}. These literature values are larger than the values measured here; we attribute this difference to increased scattering due to impurities and defects introduced during the nanopatterning process. These defects may also contribute to the separation of device and `contact lead' superconducting transitions. This separation is inconsistent even between devices on the same substrate, as demonstrated in Fig.~\ref{fig:wires}(b), which we cannot currently explain. If FIB milling were accountable for this transition separation, the separation would scale with nanopillar area. However, as shown in the inset to Fig.~\ref{fig:wires}(b), the transition temperature of the device transition, represented by the starting temperature $T_\text{device}$, does not demonstrate any dependence on device area. This uncertainty around the double transition does not affect the main observation of this paper; the appearance of the superconducting spin valve effect in both $R(H)$ and $R(T)$ measurements and a thickness dependent crossover between them. \par
The comparison of positive magnetoresistance data above and below $T_\text{c}$ supports the suggestion that CPP devices demonstrate QP GMR, as originally demonstrated in \cite{Gu2002a}. Like in that paper, we also observe a decrease in spin decay length in the superconducting state which is consistent with the additional impact of Andreev reflection causing decay of quasiparticles that cause this magnetoresistance. Unlike \cite{Gu2002a}, we have also measured superconducting devices with $d_\text{Nb}$ smaller than the thickness at which Andreev reflection does not appear to play a role. These devices are those that demonstrate the superconducting spin-valve effect, having negative $A\Delta R$. The crossover between these two effects appears to occur at around $d_\text{Nb} = 26$~nm, which is close to twice $l_\text{sf}^\text{S} = 12\pm4$~nm, similar to \cite{Gu2002a}, and further supported by Fig.~\ref{fig:dnbbyxi}, which suggests the crossover occurs where $d_\text{Nb} = 2\xi_\text{S,d}$.\par

\subsection{Phenomenological model}
A model in \cite{Gu2002a} assumes QP GMR in CPP F/S/F spin-valves decays due to Andreev reflection as quasiparticles pass a potential barrier with a temperature-dependent height i.e. $\Delta^{AP} = \Delta^P = \Delta(T)$, and reduced thickness $d_\text{Nb} - d_0$, where $d_0 / 2$ is the thickness of the region near each F/S interface where the superconducting gap and Andreev reflection are suppressed, expected to be equal to the coherence length. This approximation captures the QP GMR results in Fig.~\ref{fig:dNb} beyond $d_\text{Nb}= 26$~nm, but cannot describe our superconducting spin-valve effect data. \par
We extend these ideas from \cite{Gu2002a}, in a toy model that simply illustrates the two competing effects within our devices, capturing the overall trend as the difference of two exponential decays
\begin{equation*}
    \Delta R = A\exp{\left[\frac{-(d_\text{Nb} - d_0)}{l_\text{S}}\right]} - B\exp{\left[\frac{-(d_\text{Nb} - d_0)}{\Lambda_\text{S}}\right],}
\end{equation*}
where the first term relates to QP GMR ($\Delta R>0$), as in the model from \cite{Gu2002a} and the second to the superconducting spin valve effect ($\Delta R<0$), with a characteristic decay $\Lambda_\text{S}$. We choose an exponential for this decay to reproduce the distinct shape of the data in Fig.~\ref{fig:dNb}. In this thinner Nb regime,  $T_\text{c}(AP) > T_\text{c}(P)$ meaning that the superconducting spin-valve effect enhances $\Delta^{AP}$ \cite{Tagirov1999,Buzdin1999}; leading to decreased resistance as a greater proportion of the Nb is superconducting. We set $d_0 = 26$~nm, the crossover thickness, causing $|A/B|= 1$, and set $l_\text{S}$ to the value $l_\text{sf}^\text{S} = 12\pm4$. We find a value for $\Lambda_\text{S} = 2.2\pm0.7$~nm as an average and deviation from orthogonal distance regression on the $A\Delta R$ and $\Delta T_\text{c}$ data, which fits both data sets. This value is very small and does not initially appear to represent a physical length; however, we note previous superconducting spin-valve investigations \cite{Moraru2006} have seen a dramatic shift from a $\Delta T_\text{c}$ of 41~mK to ``only a few mK'' for an increase in $d_\text{Nb}$ of only 1~nm, which could only be modelled by theory using parameters that did not match the estimates of the authors. These short decay lengths may reflect some experimental effects not accounted for by theory based fits. 
Overall, this toy model illustrates coexistence of the superconducting spin-valve effect and QP GMR, which are competing effects within these devices. \par
Previous studies showing both positive and negative values of $\Delta R$ are explained on the basis of stray magnetic fields or vortex flow from a multi-domain state of one or both coupled F layers \cite{Zhu2009,Hwang2010,Hwang2012}. In our spin-valves positive $\Delta R$ is expected due to QP GMR, and a multidomain state with out-of-plane stray magnetic fields would cause positive magnetoresistance within the transition region, which is far from the measurement temperature of our QP GMR data [Fig.~\ref{fig:RTandRH}(b)]. No studies that we are aware of have demonstrated a systematic change of sign of magnetoresistance with $d_\text{Nb}$, such as the crossover we show here. 

\section{Conclusion}
In summary, we have presented evidence for a competition between QP GMR and the superconducting spin valve effect in superconducting CPP spin-valves. Below a Nb thickness of $d_\text{Nb} \approx 26$~nm, $\Delta R$ is negative and determined by the superconducting spin valve effect; beyond this critical thickness, $\Delta R$ is positive with a magnitude that is determined by QP Andeev reflection. This thickness appears to correspond to twice the dirty limit coherence length of the devices. These results are relevant to development of quasiparticle spintronics devices, suggesting devices utilising QP GMR effects should be fabricated with $d_\text{Nb} > 26$~nm, or consider using the superconducting spin valve effect.

\begin{acknowledgments}
This work was funded by the EPSRC through a DTP Studentship (nos. EP/M508007/1 and EP/N509620/1) and Programme Grant (no. EP/N017242/1).

\end{acknowledgments}

\appendix

\bibliography{References/bibliography.bib,References/note.bib}

\begin{thebibliography}{54}%
\makeatletter
\providecommand \@ifxundefined [1]{%
 \@ifx{#1\undefined}
}%
\providecommand \@ifnum [1]{%
 \ifnum #1\expandafter \@firstoftwo
 \else \expandafter \@secondoftwo
 \fi
}%
\providecommand \@ifx [1]{%
 \ifx #1\expandafter \@firstoftwo
 \else \expandafter \@secondoftwo
 \fi
}%
\providecommand \natexlab [1]{#1}%
\providecommand \enquote  [1]{``#1''}%
\providecommand \bibnamefont  [1]{#1}%
\providecommand \bibfnamefont [1]{#1}%
\providecommand \citenamefont [1]{#1}%
\providecommand \href@noop [0]{\@secondoftwo}%
\providecommand \href [0]{\begingroup \@sanitize@url \@href}%
\providecommand \@href[1]{\@@startlink{#1}\@@href}%
\providecommand \@@href[1]{\endgroup#1\@@endlink}%
\providecommand \@sanitize@url [0]{\catcode `\\12\catcode `\$12\catcode
  `\&12\catcode `\#12\catcode `\^12\catcode `\_12\catcode `\%12\relax}%
\providecommand \@@startlink[1]{}%
\providecommand \@@endlink[0]{}%
\providecommand \url  [0]{\begingroup\@sanitize@url \@url }%
\providecommand \@url [1]{\endgroup\@href {#1}{\urlprefix }}%
\providecommand \urlprefix  [0]{URL }%
\providecommand \Eprint [0]{\href }%
\providecommand \doibase [0]{https://doi.org/}%
\providecommand \selectlanguage [0]{\@gobble}%
\providecommand \bibinfo  [0]{\@secondoftwo}%
\providecommand \bibfield  [0]{\@secondoftwo}%
\providecommand \translation [1]{[#1]}%
\providecommand \BibitemOpen [0]{}%
\providecommand \bibitemStop [0]{}%
\providecommand \bibitemNoStop [0]{.\EOS\space}%
\providecommand \EOS [0]{\spacefactor3000\relax}%
\providecommand \BibitemShut  [1]{\csname bibitem#1\endcsname}%
\let\auto@bib@innerbib\@empty
\bibitem [{\citenamefont {{\v{Z}}uti{\`{c}}}\ \emph {et~al.}(2004)\citenamefont
  {{\v{Z}}uti{\`{c}}}, \citenamefont {Fabian},\ and\ \citenamefont {{Das
  Sarma}}}]{Zutic2004}%
  \BibitemOpen
  \bibfield  {author} {\bibinfo {author} {\bibfnamefont {I.}~\bibnamefont
  {{\v{Z}}uti{\`{c}}}}, \bibinfo {author} {\bibfnamefont {J.}~\bibnamefont
  {Fabian}},\ and\ \bibinfo {author} {\bibfnamefont {S.}~\bibnamefont {{Das
  Sarma}}},\ }\href {https://doi.org/10.1103/RevModPhys.76.323} {\bibfield
  {journal} {\bibinfo  {journal} {Rev. Mod. Phys.}\ }\textbf {\bibinfo {volume}
  {76}},\ \bibinfo {pages} {323} (\bibinfo {year} {2004})}\BibitemShut
  {NoStop}%
\bibitem [{\citenamefont {Johnson}\ and\ \citenamefont
  {Silsbee}(1985)}]{Johnson1985}%
  \BibitemOpen
  \bibfield  {author} {\bibinfo {author} {\bibfnamefont {M.}~\bibnamefont
  {Johnson}}\ and\ \bibinfo {author} {\bibfnamefont {R.~H.}\ \bibnamefont
  {Silsbee}},\ }\href {https://doi.org/10.1103/PhysRevLett.55.1790} {\bibfield
  {journal} {\bibinfo  {journal} {Phys. Rev. Lett.}\ }\textbf {\bibinfo
  {volume} {55}},\ \bibinfo {pages} {1790} (\bibinfo {year}
  {1985})}\BibitemShut {NoStop}%
\bibitem [{\citenamefont {Dieny}\ \emph {et~al.}(1991)\citenamefont {Dieny},
  \citenamefont {Speriosu}, \citenamefont {Parkin}, \citenamefont {Gurney},
  \citenamefont {Wilhoit},\ and\ \citenamefont {Mauri}}]{Dieny1991}%
  \BibitemOpen
  \bibfield  {author} {\bibinfo {author} {\bibfnamefont {B.}~\bibnamefont
  {Dieny}}, \bibinfo {author} {\bibfnamefont {V.~S.}\ \bibnamefont {Speriosu}},
  \bibinfo {author} {\bibfnamefont {S.~S.~P.}\ \bibnamefont {Parkin}}, \bibinfo
  {author} {\bibfnamefont {B.~A.}\ \bibnamefont {Gurney}}, \bibinfo {author}
  {\bibfnamefont {D.~R.}\ \bibnamefont {Wilhoit}},\ and\ \bibinfo {author}
  {\bibfnamefont {D.}~\bibnamefont {Mauri}},\ }\href
  {https://doi.org/10.1103/PhysRevB.43.1297} {\bibfield  {journal} {\bibinfo
  {journal} {Phys. Rev. B}\ }\textbf {\bibinfo {volume} {43}},\ \bibinfo
  {pages} {1297} (\bibinfo {year} {1991})}\BibitemShut {NoStop}%
\bibitem [{\citenamefont {Valet}\ and\ \citenamefont {Fert}(1993)}]{Valet1993}%
  \BibitemOpen
  \bibfield  {author} {\bibinfo {author} {\bibfnamefont {T.}~\bibnamefont
  {Valet}}\ and\ \bibinfo {author} {\bibfnamefont {A.}~\bibnamefont {Fert}},\
  }\href {https://doi.org/10.1016/0304-8853(93)91225-V} {\bibfield  {journal}
  {\bibinfo  {journal} {J. Magn. Magn. Mater.}\ }\textbf {\bibinfo {volume}
  {121}},\ \bibinfo {pages} {378} (\bibinfo {year} {1993})}\BibitemShut
  {NoStop}%
\bibitem [{\citenamefont {Bass}(2016)}]{Bass2016}%
  \BibitemOpen
  \bibfield  {author} {\bibinfo {author} {\bibfnamefont {J.}~\bibnamefont
  {Bass}},\ }\href {https://doi.org/10.1016/j.jmmm.2015.12.011} {\bibfield
  {journal} {\bibinfo  {journal} {J. Magn. Magn. Mater.}\ }\textbf {\bibinfo
  {volume} {408}},\ \bibinfo {pages} {244} (\bibinfo {year}
  {2016})}\BibitemShut {NoStop}%
\bibitem [{\citenamefont {Oh}\ \emph {et~al.}(1997)\citenamefont {Oh},
  \citenamefont {Youm},\ and\ \citenamefont {Beasley}}]{Oh1997}%
  \BibitemOpen
  \bibfield  {author} {\bibinfo {author} {\bibfnamefont {S.}~\bibnamefont
  {Oh}}, \bibinfo {author} {\bibfnamefont {D.}~\bibnamefont {Youm}},\ and\
  \bibinfo {author} {\bibfnamefont {M.~R.}\ \bibnamefont {Beasley}},\ }\href
  {https://doi.org/doi:10.1063/1.120032} {\bibfield  {journal} {\bibinfo
  {journal} {Appl. Phys. Lett.}\ }\textbf {\bibinfo {volume} {71}},\ \bibinfo
  {pages} {2376} (\bibinfo {year} {1997})}\BibitemShut {NoStop}%
\bibitem [{\citenamefont {Tagirov}(1999)}]{Tagirov1999}%
  \BibitemOpen
  \bibfield  {author} {\bibinfo {author} {\bibfnamefont {L.~R.}\ \bibnamefont
  {Tagirov}},\ }\href {https://doi.org/10.1103/PhysRevLett.83.2058} {\bibfield
  {journal} {\bibinfo  {journal} {Phys. Rev. Lett.}\ }\textbf {\bibinfo
  {volume} {83}},\ \bibinfo {pages} {2058} (\bibinfo {year}
  {1999})}\BibitemShut {NoStop}%
\bibitem [{\citenamefont {Buzdin}\ and\ \citenamefont
  {Vedyayev}(1999)}]{Buzdin1999}%
  \BibitemOpen
  \bibfield  {author} {\bibinfo {author} {\bibfnamefont {A.}~\bibnamefont
  {Buzdin}}\ and\ \bibinfo {author} {\bibfnamefont {A.}~\bibnamefont
  {Vedyayev}},\ }\href {http://iopscience.iop.org/0295-5075/48/6/686}
  {\bibfield  {journal} {\bibinfo  {journal} {Europhysics Lett.}\ }\textbf
  {\bibinfo {volume} {48}},\ \bibinfo {pages} {686} (\bibinfo {year}
  {1999})}\BibitemShut {NoStop}%
\bibitem [{\citenamefont {Gu}\ \emph {et~al.}(2002{\natexlab{a}})\citenamefont
  {Gu}, \citenamefont {You}, \citenamefont {Jiang}, \citenamefont {Pearson},
  \citenamefont {Bazaliy},\ and\ \citenamefont {Bader}}]{Gu2002}%
  \BibitemOpen
  \bibfield  {author} {\bibinfo {author} {\bibfnamefont {J.~Y.}\ \bibnamefont
  {Gu}}, \bibinfo {author} {\bibfnamefont {C.-Y.}\ \bibnamefont {You}},
  \bibinfo {author} {\bibfnamefont {J.~S.}\ \bibnamefont {Jiang}}, \bibinfo
  {author} {\bibfnamefont {J.}~\bibnamefont {Pearson}}, \bibinfo {author}
  {\bibfnamefont {Y.~B.}\ \bibnamefont {Bazaliy}},\ and\ \bibinfo {author}
  {\bibfnamefont {S.~D.}\ \bibnamefont {Bader}},\ }\href
  {https://doi.org/10.1103/PhysRevLett.89.267001} {\bibfield  {journal}
  {\bibinfo  {journal} {Phys. Rev. Lett.}\ }\textbf {\bibinfo {volume} {89}},\
  \bibinfo {pages} {267001} (\bibinfo {year} {2002}{\natexlab{a}})}\BibitemShut
  {NoStop}%
\bibitem [{\citenamefont {Potenza}\ and\ \citenamefont
  {Marrows}(2005)}]{Potenza2005}%
  \BibitemOpen
  \bibfield  {author} {\bibinfo {author} {\bibfnamefont {A.}~\bibnamefont
  {Potenza}}\ and\ \bibinfo {author} {\bibfnamefont {C.~H.}\ \bibnamefont
  {Marrows}},\ }\href {https://doi.org/10.1103/PhysRevB.71.180503} {\bibfield
  {journal} {\bibinfo  {journal} {Phys. Rev. B}\ }\textbf {\bibinfo {volume}
  {71}},\ \bibinfo {pages} {180503(R)} (\bibinfo {year} {2005})}\BibitemShut
  {NoStop}%
\bibitem [{\citenamefont {Moraru}\ \emph
  {et~al.}(2006{\natexlab{a}})\citenamefont {Moraru}, \citenamefont {Pratt},\
  and\ \citenamefont {Birge}}]{Moraru2006}%
  \BibitemOpen
  \bibfield  {author} {\bibinfo {author} {\bibfnamefont {I.~C.}\ \bibnamefont
  {Moraru}}, \bibinfo {author} {\bibfnamefont {W.~P.}\ \bibnamefont {Pratt}},\
  and\ \bibinfo {author} {\bibfnamefont {N.~O.}\ \bibnamefont {Birge}},\ }\href
  {https://doi.org/10.1103/PhysRevLett.96.037004} {\bibfield  {journal}
  {\bibinfo  {journal} {Phys. Rev. Lett.}\ }\textbf {\bibinfo {volume} {96}},\
  \bibinfo {pages} {037004} (\bibinfo {year} {2006}{\natexlab{a}})}\BibitemShut
  {NoStop}%
\bibitem [{\citenamefont {Moraru}\ \emph
  {et~al.}(2006{\natexlab{b}})\citenamefont {Moraru}, \citenamefont {Pratt},\
  and\ \citenamefont {Birge}}]{Moraru2006a}%
  \BibitemOpen
  \bibfield  {author} {\bibinfo {author} {\bibfnamefont {I.~C.}\ \bibnamefont
  {Moraru}}, \bibinfo {author} {\bibfnamefont {W.~P.}\ \bibnamefont {Pratt}},\
  and\ \bibinfo {author} {\bibfnamefont {N.~O.}\ \bibnamefont {Birge}},\ }\href
  {https://doi.org/10.1103/PhysRevB.74.220507} {\bibfield  {journal} {\bibinfo
  {journal} {Phys. Rev. B}\ }\textbf {\bibinfo {volume} {74}},\ \bibinfo
  {pages} {220507(R)} (\bibinfo {year} {2006}{\natexlab{b}})}\BibitemShut
  {NoStop}%
\bibitem [{\citenamefont {Miao}\ \emph {et~al.}(2008)\citenamefont {Miao},
  \citenamefont {Ramos},\ and\ \citenamefont {Moodera}}]{Miao2008}%
  \BibitemOpen
  \bibfield  {author} {\bibinfo {author} {\bibfnamefont {G.-X.}\ \bibnamefont
  {Miao}}, \bibinfo {author} {\bibfnamefont {A.~V.}\ \bibnamefont {Ramos}},\
  and\ \bibinfo {author} {\bibfnamefont {J.~S.}\ \bibnamefont {Moodera}},\
  }\href {https://doi.org/10.1103/PhysRevLett.101.137001} {\bibfield  {journal}
  {\bibinfo  {journal} {Phys. Rev. Lett.}\ }\textbf {\bibinfo {volume} {101}},\
  \bibinfo {pages} {137001} (\bibinfo {year} {2008})}\BibitemShut {NoStop}%
\bibitem [{\citenamefont {Zhu}\ \emph {et~al.}(2009)\citenamefont {Zhu},
  \citenamefont {Cheng}, \citenamefont {Boone},\ and\ \citenamefont
  {Krivorotov}}]{Zhu2009}%
  \BibitemOpen
  \bibfield  {author} {\bibinfo {author} {\bibfnamefont {J.}~\bibnamefont
  {Zhu}}, \bibinfo {author} {\bibfnamefont {X.}~\bibnamefont {Cheng}}, \bibinfo
  {author} {\bibfnamefont {C.}~\bibnamefont {Boone}},\ and\ \bibinfo {author}
  {\bibfnamefont {I.~N.}\ \bibnamefont {Krivorotov}},\ }\href
  {https://doi.org/10.1103/PhysRevLett.103.027004} {\bibfield  {journal}
  {\bibinfo  {journal} {Phys. Rev. Lett.}\ }\textbf {\bibinfo {volume} {103}},\
  \bibinfo {pages} {027004} (\bibinfo {year} {2009})}\BibitemShut {NoStop}%
\bibitem [{\citenamefont {Leksin}\ \emph {et~al.}(2010)\citenamefont {Leksin},
  \citenamefont {Garif'yanov}, \citenamefont {Garifullin}, \citenamefont
  {Schumann}, \citenamefont {Vinzelberg}, \citenamefont {Kataev}, \citenamefont
  {Klingeler}, \citenamefont {Schmidt},\ and\ \citenamefont
  {B{\"{u}}chner}}]{Leksin2010}%
  \BibitemOpen
  \bibfield  {author} {\bibinfo {author} {\bibfnamefont {P.~V.}\ \bibnamefont
  {Leksin}}, \bibinfo {author} {\bibfnamefont {N.~N.}\ \bibnamefont
  {Garif'yanov}}, \bibinfo {author} {\bibfnamefont {I.~A.}\ \bibnamefont
  {Garifullin}}, \bibinfo {author} {\bibfnamefont {J.}~\bibnamefont
  {Schumann}}, \bibinfo {author} {\bibfnamefont {H.}~\bibnamefont
  {Vinzelberg}}, \bibinfo {author} {\bibfnamefont {V.}~\bibnamefont {Kataev}},
  \bibinfo {author} {\bibfnamefont {R.}~\bibnamefont {Klingeler}}, \bibinfo
  {author} {\bibfnamefont {O.~G.}\ \bibnamefont {Schmidt}},\ and\ \bibinfo
  {author} {\bibfnamefont {B.}~\bibnamefont {B{\"{u}}chner}},\ }\href
  {https://doi.org/10.1063/1.3486687} {\bibfield  {journal} {\bibinfo
  {journal} {Appl. Phys. Lett.}\ }\textbf {\bibinfo {volume} {97}},\ \bibinfo
  {pages} {102505} (\bibinfo {year} {2010})}\BibitemShut {NoStop}%
\bibitem [{\citenamefont {Zhu}\ \emph {et~al.}(2010)\citenamefont {Zhu},
  \citenamefont {Krivorotov}, \citenamefont {Halterman},\ and\ \citenamefont
  {Valls}}]{Zhu2010}%
  \BibitemOpen
  \bibfield  {author} {\bibinfo {author} {\bibfnamefont {J.}~\bibnamefont
  {Zhu}}, \bibinfo {author} {\bibfnamefont {I.~N.}\ \bibnamefont {Krivorotov}},
  \bibinfo {author} {\bibfnamefont {K.}~\bibnamefont {Halterman}},\ and\
  \bibinfo {author} {\bibfnamefont {O.~T.}\ \bibnamefont {Valls}},\ }\href
  {https://doi.org/10.1103/PhysRevLett.105.207002} {\bibfield  {journal}
  {\bibinfo  {journal} {Phys. Rev. Lett.}\ }\textbf {\bibinfo {volume} {105}},\
  \bibinfo {pages} {207002} (\bibinfo {year} {2010})}\BibitemShut {NoStop}%
\bibitem [{\citenamefont {Leksin}\ \emph {et~al.}(2015)\citenamefont {Leksin},
  \citenamefont {Garif'yanov}, \citenamefont {Kamashev}, \citenamefont
  {Fominov}, \citenamefont {Schumann}, \citenamefont {Hess}, \citenamefont
  {Kataev}, \citenamefont {B{\"{u}}chner},\ and\ \citenamefont
  {Garifullin}}]{Leksin2015}%
  \BibitemOpen
  \bibfield  {author} {\bibinfo {author} {\bibfnamefont {P.~V.}\ \bibnamefont
  {Leksin}}, \bibinfo {author} {\bibfnamefont {N.~N.}\ \bibnamefont
  {Garif'yanov}}, \bibinfo {author} {\bibfnamefont {A.~A.}\ \bibnamefont
  {Kamashev}}, \bibinfo {author} {\bibfnamefont {Y.~V.}\ \bibnamefont
  {Fominov}}, \bibinfo {author} {\bibfnamefont {J.}~\bibnamefont {Schumann}},
  \bibinfo {author} {\bibfnamefont {C.}~\bibnamefont {Hess}}, \bibinfo {author}
  {\bibfnamefont {V.}~\bibnamefont {Kataev}}, \bibinfo {author} {\bibfnamefont
  {B.}~\bibnamefont {B{\"{u}}chner}},\ and\ \bibinfo {author} {\bibfnamefont
  {I.~a.}\ \bibnamefont {Garifullin}},\ }\href
  {https://doi.org/10.1103/PhysRevB.91.214508} {\bibfield  {journal} {\bibinfo
  {journal} {Phys. Rev. B}\ }\textbf {\bibinfo {volume} {91}},\ \bibinfo
  {pages} {214508} (\bibinfo {year} {2015})}\BibitemShut {NoStop}%
\bibitem [{\citenamefont {Jara}\ \emph {et~al.}(2019)\citenamefont {Jara},
  \citenamefont {Moen}, \citenamefont {Valls},\ and\ \citenamefont
  {Krivorotov}}]{Jara2019}%
  \BibitemOpen
  \bibfield  {author} {\bibinfo {author} {\bibfnamefont {A.~A.}\ \bibnamefont
  {Jara}}, \bibinfo {author} {\bibfnamefont {E.}~\bibnamefont {Moen}}, \bibinfo
  {author} {\bibfnamefont {O.~T.}\ \bibnamefont {Valls}},\ and\ \bibinfo
  {author} {\bibfnamefont {I.~N.}\ \bibnamefont {Krivorotov}},\ }\href
  {https://doi.org/10.1103/PhysRevB.100.184512} {\bibfield  {journal} {\bibinfo
   {journal} {Phys. Rev. B}\ }\textbf {\bibinfo {volume} {100}},\ \bibinfo
  {pages} {184512} (\bibinfo {year} {2019})}\BibitemShut {NoStop}%
\bibitem [{\citenamefont {Gu}\ \emph {et~al.}(2015)\citenamefont {Gu},
  \citenamefont {Hal{\'{a}}sz}, \citenamefont {Robinson},\ and\ \citenamefont
  {Blamire}}]{Gu2015}%
  \BibitemOpen
  \bibfield  {author} {\bibinfo {author} {\bibfnamefont {Y.}~\bibnamefont
  {Gu}}, \bibinfo {author} {\bibfnamefont {G.~B.}\ \bibnamefont
  {Hal{\'{a}}sz}}, \bibinfo {author} {\bibfnamefont {J.~W.~A.}\ \bibnamefont
  {Robinson}},\ and\ \bibinfo {author} {\bibfnamefont {M.~G.}\ \bibnamefont
  {Blamire}},\ }\href {https://doi.org/10.1103/PhysRevLett.115.067201}
  {\bibfield  {journal} {\bibinfo  {journal} {Phys. Rev. Lett.}\ }\textbf
  {\bibinfo {volume} {115}},\ \bibinfo {pages} {067201} (\bibinfo {year}
  {2015})}\BibitemShut {NoStop}%
\bibitem [{\citenamefont {Li}\ \emph {et~al.}(2013)\citenamefont {Li},
  \citenamefont {Roschewsky}, \citenamefont {Assaf}, \citenamefont {Eich},
  \citenamefont {Epstein-Martin}, \citenamefont {Heiman}, \citenamefont
  {M{\"{u}}nzenberg},\ and\ \citenamefont {Moodera}}]{Li2013}%
  \BibitemOpen
  \bibfield  {author} {\bibinfo {author} {\bibfnamefont {B.}~\bibnamefont
  {Li}}, \bibinfo {author} {\bibfnamefont {N.}~\bibnamefont {Roschewsky}},
  \bibinfo {author} {\bibfnamefont {B.~A.}\ \bibnamefont {Assaf}}, \bibinfo
  {author} {\bibfnamefont {M.}~\bibnamefont {Eich}}, \bibinfo {author}
  {\bibfnamefont {M.}~\bibnamefont {Epstein-Martin}}, \bibinfo {author}
  {\bibfnamefont {D.}~\bibnamefont {Heiman}}, \bibinfo {author} {\bibfnamefont
  {M.}~\bibnamefont {M{\"{u}}nzenberg}},\ and\ \bibinfo {author} {\bibfnamefont
  {J.~S.}\ \bibnamefont {Moodera}},\ }\href
  {https://doi.org/10.1103/PhysRevLett.110.097001} {\bibfield  {journal}
  {\bibinfo  {journal} {Phys. Rev. Lett.}\ }\textbf {\bibinfo {volume} {110}},\
  \bibinfo {pages} {097001} (\bibinfo {year} {2013})}\BibitemShut {NoStop}%
\bibitem [{\citenamefont {Zhu}\ \emph {et~al.}(2016)\citenamefont {Zhu},
  \citenamefont {Pal}, \citenamefont {Blamire},\ and\ \citenamefont
  {Barber}}]{Zhu2016}%
  \BibitemOpen
  \bibfield  {author} {\bibinfo {author} {\bibfnamefont {Y.}~\bibnamefont
  {Zhu}}, \bibinfo {author} {\bibfnamefont {A.}~\bibnamefont {Pal}}, \bibinfo
  {author} {\bibfnamefont {M.~G.}\ \bibnamefont {Blamire}},\ and\ \bibinfo
  {author} {\bibfnamefont {Z.~H.}\ \bibnamefont {Barber}},\ }\href
  {https://doi.org/10.1038/nmat4753} {\bibfield  {journal} {\bibinfo  {journal}
  {Nat. Mater.}\ }\textbf {\bibinfo {volume} {1}},\ \bibinfo {pages} {1}
  (\bibinfo {year} {2016})}\BibitemShut {NoStop}%
\bibitem [{\citenamefont {Aarts}\ and\ \citenamefont
  {Rusanov}(2006)}]{Aarts2006}%
  \BibitemOpen
  \bibfield  {author} {\bibinfo {author} {\bibfnamefont {J.}~\bibnamefont
  {Aarts}}\ and\ \bibinfo {author} {\bibfnamefont {A.~Y.}\ \bibnamefont
  {Rusanov}},\ }\href {https://doi.org/10.1016/j.crhy.2005.12.005} {\bibfield
  {journal} {\bibinfo  {journal} {Comptes Rendus Phys.}\ }\textbf {\bibinfo
  {volume} {7}},\ \bibinfo {pages} {99} (\bibinfo {year} {2006})}\BibitemShut
  {NoStop}%
\bibitem [{\citenamefont {Rusanov}\ \emph {et~al.}(2006)\citenamefont
  {Rusanov}, \citenamefont {Habraken},\ and\ \citenamefont
  {Aarts}}]{Rusanov2006}%
  \BibitemOpen
  \bibfield  {author} {\bibinfo {author} {\bibfnamefont {A.~Y.}\ \bibnamefont
  {Rusanov}}, \bibinfo {author} {\bibfnamefont {S.}~\bibnamefont {Habraken}},\
  and\ \bibinfo {author} {\bibfnamefont {J.}~\bibnamefont {Aarts}},\ }\href
  {https://doi.org/10.1103/PhysRevB.73.060505} {\bibfield  {journal} {\bibinfo
  {journal} {Phys. Rev. B}\ }\textbf {\bibinfo {volume} {73}},\ \bibinfo
  {pages} {060505(R)} (\bibinfo {year} {2006})}\BibitemShut {NoStop}%
\bibitem [{\citenamefont {Steiner}\ and\ \citenamefont
  {Ziemann}(2006)}]{Steiner2006}%
  \BibitemOpen
  \bibfield  {author} {\bibinfo {author} {\bibfnamefont {R.}~\bibnamefont
  {Steiner}}\ and\ \bibinfo {author} {\bibfnamefont {P.}~\bibnamefont
  {Ziemann}},\ }\href {https://doi.org/10.1103/PhysRevB.74.094504} {\bibfield
  {journal} {\bibinfo  {journal} {Phys. Rev. B}\ }\textbf {\bibinfo {volume}
  {74}},\ \bibinfo {pages} {094504} (\bibinfo {year} {2006})}\BibitemShut
  {NoStop}%
\bibitem [{\citenamefont {Singh}\ \emph
  {et~al.}(2007{\natexlab{a}})\citenamefont {Singh}, \citenamefont
  {S\"urgers},\ and\ \citenamefont {L\"ohneysen}}]{Singh2007}%
  \BibitemOpen
  \bibfield  {author} {\bibinfo {author} {\bibfnamefont {A.}~\bibnamefont
  {Singh}}, \bibinfo {author} {\bibfnamefont {C.}~\bibnamefont {S\"urgers}},\
  and\ \bibinfo {author} {\bibfnamefont {H.~v.}\ \bibnamefont {L\"ohneysen}},\
  }\href {https://doi.org/10.1103/PhysRevB.75.024513} {\bibfield  {journal}
  {\bibinfo  {journal} {Phys. Rev. B}\ }\textbf {\bibinfo {volume} {75}},\
  \bibinfo {pages} {024513} (\bibinfo {year} {2007}{\natexlab{a}})}\BibitemShut
  {NoStop}%
\bibitem [{\citenamefont {Singh}\ \emph
  {et~al.}(2007{\natexlab{b}})\citenamefont {Singh}, \citenamefont
  {S{\"{u}}rgers}, \citenamefont {Hoffmann}, \citenamefont {L{\"{o}}hneysen},
  \citenamefont {Ashworth}, \citenamefont {Pilet},\ and\ \citenamefont
  {Hug}}]{Singh2007a}%
  \BibitemOpen
  \bibfield  {author} {\bibinfo {author} {\bibfnamefont {A.}~\bibnamefont
  {Singh}}, \bibinfo {author} {\bibfnamefont {C.}~\bibnamefont
  {S{\"{u}}rgers}}, \bibinfo {author} {\bibfnamefont {R.}~\bibnamefont
  {Hoffmann}}, \bibinfo {author} {\bibfnamefont {H.~V.}\ \bibnamefont
  {L{\"{o}}hneysen}}, \bibinfo {author} {\bibfnamefont {T.~V.}\ \bibnamefont
  {Ashworth}}, \bibinfo {author} {\bibfnamefont {N.}~\bibnamefont {Pilet}},\
  and\ \bibinfo {author} {\bibfnamefont {H.~J.}\ \bibnamefont {Hug}},\ }\href
  {https://doi.org/10.1063/1.2794424} {\bibfield  {journal} {\bibinfo
  {journal} {Appl. Phys. Lett.}\ }\textbf {\bibinfo {volume} {91}},\ \bibinfo
  {pages} {71} (\bibinfo {year} {2007}{\natexlab{b}})}\BibitemShut {NoStop}%
\bibitem [{\citenamefont {Stamopoulos}\ \emph {et~al.}(2007)\citenamefont
  {Stamopoulos}, \citenamefont {Manios},\ and\ \citenamefont
  {Pissas}}]{Stamopoulos2007a}%
  \BibitemOpen
  \bibfield  {author} {\bibinfo {author} {\bibfnamefont {D.}~\bibnamefont
  {Stamopoulos}}, \bibinfo {author} {\bibfnamefont {E.}~\bibnamefont
  {Manios}},\ and\ \bibinfo {author} {\bibfnamefont {M.}~\bibnamefont
  {Pissas}},\ }\href {https://doi.org/10.1103/PhysRevB.75.184504} {\bibfield
  {journal} {\bibinfo  {journal} {Phys. Rev. B}\ }\textbf {\bibinfo {volume}
  {75}},\ \bibinfo {pages} {184504} (\bibinfo {year} {2007})}\BibitemShut
  {NoStop}%
\bibitem [{\citenamefont {Leksin}\ \emph {et~al.}(2009)\citenamefont {Leksin},
  \citenamefont {Salikhov}, \citenamefont {Garifullin}, \citenamefont
  {Vinzelberg}, \citenamefont {Kataev}, \citenamefont {Klingeler},
  \citenamefont {Tagirov},\ and\ \citenamefont {B{\"{u}}chner}}]{Leksin2009}%
  \BibitemOpen
  \bibfield  {author} {\bibinfo {author} {\bibfnamefont {P.~V.}\ \bibnamefont
  {Leksin}}, \bibinfo {author} {\bibfnamefont {R.~I.}\ \bibnamefont
  {Salikhov}}, \bibinfo {author} {\bibfnamefont {I.~A.}\ \bibnamefont
  {Garifullin}}, \bibinfo {author} {\bibfnamefont {H.}~\bibnamefont
  {Vinzelberg}}, \bibinfo {author} {\bibfnamefont {V.}~\bibnamefont {Kataev}},
  \bibinfo {author} {\bibfnamefont {R.}~\bibnamefont {Klingeler}}, \bibinfo
  {author} {\bibfnamefont {L.~R.}\ \bibnamefont {Tagirov}},\ and\ \bibinfo
  {author} {\bibfnamefont {B.}~\bibnamefont {B{\"{u}}chner}},\ }\href
  {https://doi.org/10.1134/S0021364009130128} {\bibfield  {journal} {\bibinfo
  {journal} {JETP Lett.}\ }\textbf {\bibinfo {volume} {90}},\ \bibinfo {pages}
  {59} (\bibinfo {year} {2009})}\BibitemShut {NoStop}%
\bibitem [{\citenamefont {Hwang}\ \emph {et~al.}(2010)\citenamefont {Hwang},
  \citenamefont {Kim},\ and\ \citenamefont {Oh}}]{Hwang2010}%
  \BibitemOpen
  \bibfield  {author} {\bibinfo {author} {\bibfnamefont {T.~J.}\ \bibnamefont
  {Hwang}}, \bibinfo {author} {\bibfnamefont {D.~H.}\ \bibnamefont {Kim}},\
  and\ \bibinfo {author} {\bibfnamefont {S.}~\bibnamefont {Oh}},\ }\href
  {https://doi.org/10.1109/TMAG.2009.2032144} {\bibfield  {journal} {\bibinfo
  {journal} {IEEE Trans. Magn.}\ }\textbf {\bibinfo {volume} {46}},\ \bibinfo
  {pages} {235} (\bibinfo {year} {2010})}\BibitemShut {NoStop}%
\bibitem [{\citenamefont {Flokstra}\ \emph {et~al.}(2010)\citenamefont
  {Flokstra}, \citenamefont {van~der Knaap},\ and\ \citenamefont
  {Aarts}}]{Flokstra2010a}%
  \BibitemOpen
  \bibfield  {author} {\bibinfo {author} {\bibfnamefont {M.}~\bibnamefont
  {Flokstra}}, \bibinfo {author} {\bibfnamefont {J.~M.}\ \bibnamefont {van~der
  Knaap}},\ and\ \bibinfo {author} {\bibfnamefont {J.}~\bibnamefont {Aarts}},\
  }\href {https://doi.org/10.1103/PhysRevB.82.184523} {\bibfield  {journal}
  {\bibinfo  {journal} {Phys. Rev. B}\ }\textbf {\bibinfo {volume} {82}},\
  \bibinfo {pages} {184523} (\bibinfo {year} {2010})}\BibitemShut {NoStop}%
\bibitem [{\citenamefont {Takahashi}\ \emph {et~al.}(1999)\citenamefont
  {Takahashi}, \citenamefont {Imamura},\ and\ \citenamefont
  {Maekawa}}]{Takahashi1999}%
  \BibitemOpen
  \bibfield  {author} {\bibinfo {author} {\bibfnamefont {S.}~\bibnamefont
  {Takahashi}}, \bibinfo {author} {\bibfnamefont {H.}~\bibnamefont {Imamura}},\
  and\ \bibinfo {author} {\bibfnamefont {S.}~\bibnamefont {Maekawa}},\ }\href
  {https://doi.org/10.1103/PhysRevLett.82.3911} {\bibfield  {journal} {\bibinfo
   {journal} {Phys. Rev. Lett.}\ }\textbf {\bibinfo {volume} {82}},\ \bibinfo
  {pages} {3911} (\bibinfo {year} {1999})}\BibitemShut {NoStop}%
\bibitem [{\citenamefont {Takahashi}\ and\ \citenamefont
  {Maekawa}(2003)}]{Takahashi2003}%
  \BibitemOpen
  \bibfield  {author} {\bibinfo {author} {\bibfnamefont {S.}~\bibnamefont
  {Takahashi}}\ and\ \bibinfo {author} {\bibfnamefont {S.}~\bibnamefont
  {Maekawa}},\ }\href {https://doi.org/10.1103/PhysRevB.67.052409} {\bibfield
  {journal} {\bibinfo  {journal} {Phys. Rev. B}\ }\textbf {\bibinfo {volume}
  {67}},\ \bibinfo {pages} {052409} (\bibinfo {year} {2003})}\BibitemShut
  {NoStop}%
\bibitem [{\citenamefont {Pratt}\ \emph {et~al.}(1991)\citenamefont {Pratt},
  \citenamefont {Lee}, \citenamefont {Slaughter}, \citenamefont {Loloee},
  \citenamefont {Schroeder},\ and\ \citenamefont {Bass}}]{Pratt1991}%
  \BibitemOpen
  \bibfield  {author} {\bibinfo {author} {\bibfnamefont {W.~P.}\ \bibnamefont
  {Pratt}}, \bibinfo {author} {\bibfnamefont {S.~F.}\ \bibnamefont {Lee}},
  \bibinfo {author} {\bibfnamefont {J.~M.}\ \bibnamefont {Slaughter}}, \bibinfo
  {author} {\bibfnamefont {R.}~\bibnamefont {Loloee}}, \bibinfo {author}
  {\bibfnamefont {P.~A.}\ \bibnamefont {Schroeder}},\ and\ \bibinfo {author}
  {\bibfnamefont {J.}~\bibnamefont {Bass}},\ }\href
  {https://doi.org/10.1103/PhysRevLett.66.3060} {\bibfield  {journal} {\bibinfo
   {journal} {Phys. Rev. Lett.}\ }\textbf {\bibinfo {volume} {66}},\ \bibinfo
  {pages} {3060} (\bibinfo {year} {1991})}\BibitemShut {NoStop}%
\bibitem [{\citenamefont {Lee}\ \emph {et~al.}(1995)\citenamefont {Lee},
  \citenamefont {Yang}, \citenamefont {Holody}, \citenamefont {Loloee},
  \citenamefont {Hetherington}, \citenamefont {Mahmood}, \citenamefont
  {Ikegami}, \citenamefont {Vigen}, \citenamefont {Henry}, \citenamefont
  {Schroeder}, \citenamefont {Pratt},\ and\ \citenamefont {Bass}}]{Lee1995}%
  \BibitemOpen
  \bibfield  {author} {\bibinfo {author} {\bibfnamefont {S.~F.}\ \bibnamefont
  {Lee}}, \bibinfo {author} {\bibfnamefont {Q.}~\bibnamefont {Yang}}, \bibinfo
  {author} {\bibfnamefont {P.}~\bibnamefont {Holody}}, \bibinfo {author}
  {\bibfnamefont {R.}~\bibnamefont {Loloee}}, \bibinfo {author} {\bibfnamefont
  {J.~H.}\ \bibnamefont {Hetherington}}, \bibinfo {author} {\bibfnamefont
  {S.}~\bibnamefont {Mahmood}}, \bibinfo {author} {\bibfnamefont
  {B.}~\bibnamefont {Ikegami}}, \bibinfo {author} {\bibfnamefont
  {K.}~\bibnamefont {Vigen}}, \bibinfo {author} {\bibfnamefont {L.~L.}\
  \bibnamefont {Henry}}, \bibinfo {author} {\bibfnamefont {P.~A.}\ \bibnamefont
  {Schroeder}}, \bibinfo {author} {\bibfnamefont {W.~P.}\ \bibnamefont
  {Pratt}},\ and\ \bibinfo {author} {\bibfnamefont {J.}~\bibnamefont {Bass}},\
  }\href {https://doi.org/10.1103/PhysRevB.52.15426} {\bibfield  {journal}
  {\bibinfo  {journal} {Phys. Rev. B}\ }\textbf {\bibinfo {volume} {52}},\
  \bibinfo {pages} {15426} (\bibinfo {year} {1995})}\BibitemShut {NoStop}%
\bibitem [{\citenamefont {Gijs}\ \emph {et~al.}(1995)\citenamefont {Gijs},
  \citenamefont {Lenczowski}, \citenamefont {Giesbers}, \citenamefont {van~de
  Veerdonk}, \citenamefont {Johnson},\ and\ \citenamefont {aan~de
  Stegge}}]{Gijs1995}%
  \BibitemOpen
  \bibfield  {author} {\bibinfo {author} {\bibfnamefont {M.~A.}\ \bibnamefont
  {Gijs}}, \bibinfo {author} {\bibfnamefont {S.~K.}\ \bibnamefont
  {Lenczowski}}, \bibinfo {author} {\bibfnamefont {J.~B.}\ \bibnamefont
  {Giesbers}}, \bibinfo {author} {\bibfnamefont {R.~J.}\ \bibnamefont {van~de
  Veerdonk}}, \bibinfo {author} {\bibfnamefont {M.~T.}\ \bibnamefont
  {Johnson}},\ and\ \bibinfo {author} {\bibfnamefont {J.~B.}\ \bibnamefont
  {aan~de Stegge}},\ }\href {https://doi.org/10.1016/0921-5107(94)08023-2}
  {\bibfield  {journal} {\bibinfo  {journal} {Mater. Sci. Eng. B}\ }\textbf
  {\bibinfo {volume} {31}},\ \bibinfo {pages} {85} (\bibinfo {year}
  {1995})}\BibitemShut {NoStop}%
\bibitem [{\citenamefont {Eid}\ \emph {et~al.}(2002)\citenamefont {Eid},
  \citenamefont {Portner}, \citenamefont {Borchers}, \citenamefont {Loloee},
  \citenamefont {{Al-Haj Darwish}}, \citenamefont {Tsoi}, \citenamefont
  {Slater}, \citenamefont {O'Donovan}, \citenamefont {Kurt}, \citenamefont
  {Pratt},\ and\ \citenamefont {Bass}}]{Eid2002}%
  \BibitemOpen
  \bibfield  {author} {\bibinfo {author} {\bibfnamefont {K.}~\bibnamefont
  {Eid}}, \bibinfo {author} {\bibfnamefont {D.}~\bibnamefont {Portner}},
  \bibinfo {author} {\bibfnamefont {J.~A.}\ \bibnamefont {Borchers}}, \bibinfo
  {author} {\bibfnamefont {R.}~\bibnamefont {Loloee}}, \bibinfo {author}
  {\bibfnamefont {M.}~\bibnamefont {{Al-Haj Darwish}}}, \bibinfo {author}
  {\bibfnamefont {M.}~\bibnamefont {Tsoi}}, \bibinfo {author} {\bibfnamefont
  {R.~D.}\ \bibnamefont {Slater}}, \bibinfo {author} {\bibfnamefont {K.~V.}\
  \bibnamefont {O'Donovan}}, \bibinfo {author} {\bibfnamefont {H.}~\bibnamefont
  {Kurt}}, \bibinfo {author} {\bibfnamefont {W.~P.}\ \bibnamefont {Pratt}},\
  and\ \bibinfo {author} {\bibfnamefont {J.}~\bibnamefont {Bass}},\ }\href
  {https://doi.org/10.1103/PhysRevB.65.054424} {\bibfield  {journal} {\bibinfo
  {journal} {Phys. Rev. B}\ }\textbf {\bibinfo {volume} {65}},\ \bibinfo
  {pages} {544241} (\bibinfo {year} {2002})}\BibitemShut {NoStop}%
\bibitem [{\citenamefont {Gu}\ \emph {et~al.}(2002{\natexlab{b}})\citenamefont
  {Gu}, \citenamefont {Caballero}, \citenamefont {Slater}, \citenamefont
  {Loloee},\ and\ \citenamefont {Pratt}}]{Gu2002a}%
  \BibitemOpen
  \bibfield  {author} {\bibinfo {author} {\bibfnamefont {J.~Y.}\ \bibnamefont
  {Gu}}, \bibinfo {author} {\bibfnamefont {J.~A.}\ \bibnamefont {Caballero}},
  \bibinfo {author} {\bibfnamefont {R.~D.}\ \bibnamefont {Slater}}, \bibinfo
  {author} {\bibfnamefont {R.}~\bibnamefont {Loloee}},\ and\ \bibinfo {author}
  {\bibfnamefont {W.~P.}\ \bibnamefont {Pratt}},\ }\href
  {https://doi.org/10.1103/PhysRevB.66.140507} {\bibfield  {journal} {\bibinfo
  {journal} {Phys. Rev. B}\ }\textbf {\bibinfo {volume} {66}},\ \bibinfo
  {pages} {140507(R)} (\bibinfo {year} {2002}{\natexlab{b}})}\BibitemShut
  {NoStop}%
\bibitem [{\citenamefont {Bell}\ \emph {et~al.}(2005)\citenamefont {Bell},
  \citenamefont {Loloee}, \citenamefont {Burnell},\ and\ \citenamefont
  {Blamire}}]{Bell2005}%
  \BibitemOpen
  \bibfield  {author} {\bibinfo {author} {\bibfnamefont {C.}~\bibnamefont
  {Bell}}, \bibinfo {author} {\bibfnamefont {R.}~\bibnamefont {Loloee}},
  \bibinfo {author} {\bibfnamefont {G.}~\bibnamefont {Burnell}},\ and\ \bibinfo
  {author} {\bibfnamefont {M.~G.}\ \bibnamefont {Blamire}},\ }\href
  {https://doi.org/10.1103/PhysRevB.71.180501} {\bibfield  {journal} {\bibinfo
  {journal} {Phys. Rev. B}\ }\textbf {\bibinfo {volume} {71}},\ \bibinfo
  {pages} {180501(R)} (\bibinfo {year} {2005})}\BibitemShut {NoStop}%
\bibitem [{\citenamefont {Robinson}\ \emph {et~al.}(2007)\citenamefont
  {Robinson}, \citenamefont {Piano}, \citenamefont {Burnell}, \citenamefont
  {Bell},\ and\ \citenamefont {Blamire}}]{Robinson2007}%
  \BibitemOpen
  \bibfield  {author} {\bibinfo {author} {\bibfnamefont {J.~W.~A.}\
  \bibnamefont {Robinson}}, \bibinfo {author} {\bibfnamefont {S.}~\bibnamefont
  {Piano}}, \bibinfo {author} {\bibfnamefont {G.}~\bibnamefont {Burnell}},
  \bibinfo {author} {\bibfnamefont {C.}~\bibnamefont {Bell}},\ and\ \bibinfo
  {author} {\bibfnamefont {M.~G.}\ \bibnamefont {Blamire}},\ }\href
  {https://doi.org/10.1103/PhysRevB.76.094522} {\bibfield  {journal} {\bibinfo
  {journal} {Phys. Rev. B}\ }\textbf {\bibinfo {volume} {76}},\ \bibinfo
  {pages} {094522} (\bibinfo {year} {2007})}\BibitemShut {NoStop}%
\bibitem [{\citenamefont {Tateishi}\ and\ \citenamefont
  {Bergmann}(2010)}]{Tateishi2010}%
  \BibitemOpen
  \bibfield  {author} {\bibinfo {author} {\bibfnamefont {G.}~\bibnamefont
  {Tateishi}}\ and\ \bibinfo {author} {\bibfnamefont {G.}~\bibnamefont
  {Bergmann}},\ }\href {https://doi.org/10.1140/epjb/e2009-00422-x} {\bibfield
  {journal} {\bibinfo  {journal} {Eur. Phys. J. B}\ }\textbf {\bibinfo {volume}
  {73}},\ \bibinfo {pages} {155} (\bibinfo {year} {2010})}\BibitemShut
  {NoStop}%
\bibitem [{\citenamefont {Bell}\ \emph {et~al.}(2003)\citenamefont {Bell},
  \citenamefont {Burnell}, \citenamefont {Kang}, \citenamefont {Hadfield},
  \citenamefont {Kappers},\ and\ \citenamefont {Blamire}}]{Bell2003a}%
  \BibitemOpen
  \bibfield  {author} {\bibinfo {author} {\bibfnamefont {C.}~\bibnamefont
  {Bell}}, \bibinfo {author} {\bibfnamefont {G.}~\bibnamefont {Burnell}},
  \bibinfo {author} {\bibfnamefont {D.-J.}\ \bibnamefont {Kang}}, \bibinfo
  {author} {\bibfnamefont {R.~H.}\ \bibnamefont {Hadfield}}, \bibinfo {author}
  {\bibfnamefont {M.~J.}\ \bibnamefont {Kappers}},\ and\ \bibinfo {author}
  {\bibfnamefont {M.~G.}\ \bibnamefont {Blamire}},\ }\href
  {https://doi.org/10.1088/0957-4484/14/6/312} {\bibfield  {journal} {\bibinfo
  {journal} {Nanotechnology}\ }\textbf {\bibinfo {volume} {14}},\ \bibinfo
  {pages} {630} (\bibinfo {year} {2003})}\BibitemShut {NoStop}%
\bibitem [{SM()}]{SM}%
  \BibitemOpen
  \href@noop {} {}\bibinfo {note} {See Supplemental Material at [URL will be
  inserted by publisher] for (I) data and additional figures, (II) the wires
  leading to the device and (III) fitting details for above and below
  $T_\text{c}$ fits.}\BibitemShut {Stop}%
\bibitem [{\citenamefont {Park}\ \emph {et~al.}(2000)\citenamefont {Park},
  \citenamefont {Baxter}, \citenamefont {Steenwyk}, \citenamefont {Moraru},
  \citenamefont {Pratt},\ and\ \citenamefont {Bass}}]{Park2000}%
  \BibitemOpen
  \bibfield  {author} {\bibinfo {author} {\bibfnamefont {W.}~\bibnamefont
  {Park}}, \bibinfo {author} {\bibfnamefont {D.~V.}\ \bibnamefont {Baxter}},
  \bibinfo {author} {\bibfnamefont {S.}~\bibnamefont {Steenwyk}}, \bibinfo
  {author} {\bibfnamefont {I.}~\bibnamefont {Moraru}}, \bibinfo {author}
  {\bibfnamefont {W.~P.}\ \bibnamefont {Pratt}},\ and\ \bibinfo {author}
  {\bibfnamefont {J.}~\bibnamefont {Bass}},\ }\href
  {https://doi.org/10.1103/PhysRevB.62.1178} {\bibfield  {journal} {\bibinfo
  {journal} {Phys. Rev. B}\ }\textbf {\bibinfo {volume} {62}},\ \bibinfo
  {pages} {1178} (\bibinfo {year} {2000})}\BibitemShut {NoStop}%
\bibitem [{\citenamefont {Vila}\ \emph {et~al.}(2000)\citenamefont {Vila},
  \citenamefont {Park}, \citenamefont {Caballero}, \citenamefont {Bozec},
  \citenamefont {Loloee}, \citenamefont {Pratt},\ and\ \citenamefont
  {Bass}}]{Vila2000}%
  \BibitemOpen
  \bibfield  {author} {\bibinfo {author} {\bibfnamefont {L.}~\bibnamefont
  {Vila}}, \bibinfo {author} {\bibfnamefont {W.}~\bibnamefont {Park}}, \bibinfo
  {author} {\bibfnamefont {J.~A.}\ \bibnamefont {Caballero}}, \bibinfo {author}
  {\bibfnamefont {D.}~\bibnamefont {Bozec}}, \bibinfo {author} {\bibfnamefont
  {R.}~\bibnamefont {Loloee}}, \bibinfo {author} {\bibfnamefont {W.~P.}\
  \bibnamefont {Pratt}},\ and\ \bibinfo {author} {\bibfnamefont
  {J.}~\bibnamefont {Bass}},\ }\href {https://doi.org/10.1063/1.373586}
  {\bibfield  {journal} {\bibinfo  {journal} {J. Appl. Phys.}\ }\textbf
  {\bibinfo {volume} {87}},\ \bibinfo {pages} {8610} (\bibinfo {year}
  {2000})}\BibitemShut {NoStop}%
\bibitem [{\citenamefont {Mcguire}\ and\ \citenamefont
  {Potter}(1975)}]{Mcguire1975}%
  \BibitemOpen
  \bibfield  {author} {\bibinfo {author} {\bibfnamefont {T.~R.}\ \bibnamefont
  {Mcguire}}\ and\ \bibinfo {author} {\bibfnamefont {R.~I.}\ \bibnamefont
  {Potter}},\ }\href {https://doi.org/10.1109/TMAG.1975.1058782} {\bibfield
  {journal} {\bibinfo  {journal} {IEEE Trans. Magn.}\ }\textbf {\bibinfo
  {volume} {11}},\ \bibinfo {pages} {1018} (\bibinfo {year}
  {1975})}\BibitemShut {NoStop}%
\bibitem [{\citenamefont {Deutscher}\ and\ \citenamefont
  {Feinberg}(2000)}]{Deutscher2000}%
  \BibitemOpen
  \bibfield  {author} {\bibinfo {author} {\bibfnamefont {G.}~\bibnamefont
  {Deutscher}}\ and\ \bibinfo {author} {\bibfnamefont {D.}~\bibnamefont
  {Feinberg}},\ }\href {https://doi.org/10.1063/1.125796} {\bibfield  {journal}
  {\bibinfo  {journal} {Appl. Phys. Lett.}\ }\textbf {\bibinfo {volume} {76}},\
  \bibinfo {pages} {487} (\bibinfo {year} {2000})}\BibitemShut {NoStop}%
\bibitem [{\citenamefont {Beckmann}\ \emph {et~al.}(2004)\citenamefont
  {Beckmann}, \citenamefont {Weber},\ and\ \citenamefont
  {L{\"{o}}hneysen}}]{Beckmann2004}%
  \BibitemOpen
  \bibfield  {author} {\bibinfo {author} {\bibfnamefont {D.}~\bibnamefont
  {Beckmann}}, \bibinfo {author} {\bibfnamefont {E.~B.}\ \bibnamefont
  {Weber}},\ and\ \bibinfo {author} {\bibfnamefont {H.~V.}\ \bibnamefont
  {L{\"{o}}hneysen}},\ }\href {https://doi.org/10.1103/PhysRevLett.93.197003}
  {\bibfield  {journal} {\bibinfo  {journal} {Phys. Rev. Lett.}\ }\textbf
  {\bibinfo {volume} {93}},\ \bibinfo {pages} {197003} (\bibinfo {year}
  {2004})}\BibitemShut {NoStop}%
\bibitem [{\citenamefont {Beckmann}\ and\ \citenamefont
  {L{\"{o}}hneysen}(2007)}]{Beckmann2007}%
  \BibitemOpen
  \bibfield  {author} {\bibinfo {author} {\bibfnamefont {D.}~\bibnamefont
  {Beckmann}}\ and\ \bibinfo {author} {\bibfnamefont {H.~V.}\ \bibnamefont
  {L{\"{o}}hneysen}},\ }\href {https://doi.org/10.1007/s00339-007-4193-4}
  {\bibfield  {journal} {\bibinfo  {journal} {Appl. Phys. A Mater. Sci.
  Process.}\ }\textbf {\bibinfo {volume} {89}},\ \bibinfo {pages} {603}
  (\bibinfo {year} {2007})}\BibitemShut {NoStop}%
\bibitem [{\citenamefont {Kleine}\ \emph {et~al.}(2009)\citenamefont {Kleine},
  \citenamefont {Baumgartner}, \citenamefont {Trbovic},\ and\ \citenamefont
  {Sch{\"{o}}nenberger}}]{Kleine2009}%
  \BibitemOpen
  \bibfield  {author} {\bibinfo {author} {\bibfnamefont {A.}~\bibnamefont
  {Kleine}}, \bibinfo {author} {\bibfnamefont {A.}~\bibnamefont {Baumgartner}},
  \bibinfo {author} {\bibfnamefont {J.}~\bibnamefont {Trbovic}},\ and\ \bibinfo
  {author} {\bibfnamefont {C.}~\bibnamefont {Sch{\"{o}}nenberger}},\ }\bibfield
   {journal} {\bibinfo  {journal} {Europhys. Lett.}\ }\textbf {\bibinfo
  {volume} {87}},\ \href {https://doi.org/10.1209/0295-5075/87/27011}
  {10.1209/0295-5075/87/27011} (\bibinfo {year} {2009})\BibitemShut {NoStop}%
\bibitem [{\citenamefont {Webb}\ \emph {et~al.}(2012)\citenamefont {Webb},
  \citenamefont {Hickey},\ and\ \citenamefont {Burnell}}]{Webb2012}%
  \BibitemOpen
  \bibfield  {author} {\bibinfo {author} {\bibfnamefont {J.~L.}\ \bibnamefont
  {Webb}}, \bibinfo {author} {\bibfnamefont {B.~J.}\ \bibnamefont {Hickey}},\
  and\ \bibinfo {author} {\bibfnamefont {G.}~\bibnamefont {Burnell}},\ }\href
  {https://doi.org/10.1103/PhysRevB.86.054525} {\bibfield  {journal} {\bibinfo
  {journal} {Phys. Rev. B}\ }\textbf {\bibinfo {volume} {86}},\ \bibinfo
  {pages} {054525} (\bibinfo {year} {2012})}\BibitemShut {NoStop}%
\bibitem [{\citenamefont {Giazotto}\ \emph {et~al.}(2006)\citenamefont
  {Giazotto}, \citenamefont {Taddei}, \citenamefont {Beltram},\ and\
  \citenamefont {Fazio}}]{Giazotto2006}%
  \BibitemOpen
  \bibfield  {author} {\bibinfo {author} {\bibfnamefont {F.}~\bibnamefont
  {Giazotto}}, \bibinfo {author} {\bibfnamefont {F.}~\bibnamefont {Taddei}},
  \bibinfo {author} {\bibfnamefont {F.}~\bibnamefont {Beltram}},\ and\ \bibinfo
  {author} {\bibfnamefont {R.}~\bibnamefont {Fazio}},\ }\href
  {https://doi.org/10.1103/PhysRevLett.97.087001} {\bibfield  {journal}
  {\bibinfo  {journal} {Phys. Rev. Lett.}\ }\textbf {\bibinfo {volume} {97}},\
  \bibinfo {pages} {087001} (\bibinfo {year} {2006})}\BibitemShut {NoStop}%
\bibitem [{\citenamefont {Cadden-Zimansky}\ \emph {et~al.}(2007)\citenamefont
  {Cadden-Zimansky}, \citenamefont {Jiang},\ and\ \citenamefont
  {Chandrasekhar}}]{Cadden-Zimansky2007}%
  \BibitemOpen
  \bibfield  {author} {\bibinfo {author} {\bibfnamefont {P.}~\bibnamefont
  {Cadden-Zimansky}}, \bibinfo {author} {\bibfnamefont {Z.}~\bibnamefont
  {Jiang}},\ and\ \bibinfo {author} {\bibfnamefont {V.}~\bibnamefont
  {Chandrasekhar}},\ }\bibfield  {journal} {\bibinfo  {journal} {New J. Phys.}\
  }\textbf {\bibinfo {volume} {9}},\ \href
  {https://doi.org/10.1088/1367-2630/9/5/116} {10.1088/1367-2630/9/5/116}
  (\bibinfo {year} {2007})\BibitemShut {NoStop}%
\bibitem [{\citenamefont {Jeon}\ \emph {et~al.}(2018)\citenamefont {Jeon},
  \citenamefont {Ciccarelli}, \citenamefont {Ferguson}, \citenamefont
  {Kurebayashi}, \citenamefont {Cohen}, \citenamefont {Montiel}, \citenamefont
  {Eschrig}, \citenamefont {Robinson},\ and\ \citenamefont
  {Blamire}}]{Jeon2018}%
  \BibitemOpen
  \bibfield  {author} {\bibinfo {author} {\bibfnamefont {K.-R.}\ \bibnamefont
  {Jeon}}, \bibinfo {author} {\bibfnamefont {C.}~\bibnamefont {Ciccarelli}},
  \bibinfo {author} {\bibfnamefont {A.~J.}\ \bibnamefont {Ferguson}}, \bibinfo
  {author} {\bibfnamefont {H.}~\bibnamefont {Kurebayashi}}, \bibinfo {author}
  {\bibfnamefont {L.~F.}\ \bibnamefont {Cohen}}, \bibinfo {author}
  {\bibfnamefont {X.}~\bibnamefont {Montiel}}, \bibinfo {author} {\bibfnamefont
  {M.}~\bibnamefont {Eschrig}}, \bibinfo {author} {\bibfnamefont {J.~W.~A.}\
  \bibnamefont {Robinson}},\ and\ \bibinfo {author} {\bibfnamefont {M.~G.}\
  \bibnamefont {Blamire}},\ }\href {https://doi.org/10.1038/s41563-018-0058-9}
  {\bibfield  {journal} {\bibinfo  {journal} {Nat. Mater.}\ }\textbf {\bibinfo
  {volume} {17}},\ \bibinfo {pages} {499} (\bibinfo {year} {2018})}\BibitemShut
  {NoStop}%
\bibitem [{\citenamefont {Hwang}\ and\ \citenamefont {Kim}(2012)}]{Hwang2012}%
  \BibitemOpen
  \bibfield  {author} {\bibinfo {author} {\bibfnamefont {T.~J.}\ \bibnamefont
  {Hwang}}\ and\ \bibinfo {author} {\bibfnamefont {D.~H.}\ \bibnamefont
  {Kim}},\ }\href {https://doi.org/10.3938/jkps.61.1628} {\bibfield  {journal}
  {\bibinfo  {journal} {J. Korean Phys. Soc.}\ }\textbf {\bibinfo {volume}
  {61}},\ \bibinfo {pages} {1628} (\bibinfo {year} {2012})}\BibitemShut
  {NoStop}%
\end{thebibliography}%


\begin{thebibliography}{0}%
\makeatletter
\providecommand \@ifxundefined [1]{%
 \@ifx{#1\undefined}
}%
\providecommand \@ifnum [1]{%
 \ifnum #1\expandafter \@firstoftwo
 \else \expandafter \@secondoftwo
 \fi
}%
\providecommand \@ifx [1]{%
 \ifx #1\expandafter \@firstoftwo
 \else \expandafter \@secondoftwo
 \fi
}%
\providecommand \natexlab [1]{#1}%
\providecommand \enquote  [1]{``#1''}%
\providecommand \bibnamefont  [1]{#1}%
\providecommand \bibfnamefont [1]{#1}%
\providecommand \citenamefont [1]{#1}%
\providecommand \href@noop [0]{\@secondoftwo}%
\providecommand \href [0]{\begingroup \@sanitize@url \@href}%
\providecommand \@href[1]{\@@startlink{#1}\@@href}%
\providecommand \@@href[1]{\endgroup#1\@@endlink}%
\providecommand \@sanitize@url [0]{\catcode `\\12\catcode `\$12\catcode
  `\&12\catcode `\#12\catcode `\^12\catcode `\_12\catcode `\%12\relax}%
\providecommand \@@startlink[1]{}%
\providecommand \@@endlink[0]{}%
\providecommand \url  [0]{\begingroup\@sanitize@url \@url }%
\providecommand \@url [1]{\endgroup\@href {#1}{\urlprefix }}%
\providecommand \urlprefix  [0]{URL }%
\providecommand \Eprint [0]{\href }%
\providecommand \doibase [0]{https://doi.org/}%
\providecommand \selectlanguage [0]{\@gobble}%
\providecommand \bibinfo  [0]{\@secondoftwo}%
\providecommand \bibfield  [0]{\@secondoftwo}%
\providecommand \translation [1]{[#1]}%
\providecommand \BibitemOpen [0]{}%
\providecommand \bibitemStop [0]{}%
\providecommand \bibitemNoStop [0]{.\EOS\space}%
\providecommand \EOS [0]{\spacefactor3000\relax}%
\providecommand \BibitemShut  [1]{\csname bibitem#1\endcsname}%
\let\auto@bib@innerbib\@empty
\end{thebibliography}%

\end{document}


\title{Evidence for competition between the superconductor proximity effect and quasiparticle spin-decay in superconducting spin-valves:\\Supplemental Material}

\author{B. Stoddart-Stones}
\email{bs507@cam.ac.uk}
\author{X. Montiel}
\author{M. G. Blamire}
\author{J. W. A. Robinson}
\email{jjr33@cam.ac.uk}
\affiliation{%
 Department of Materials Science \& Metallurgy, University of Cambridge, 27 Charles Babbage Road, Cambridge CB3 0FS, United Kingdom
}%

\date{\today}

\maketitle
\beginsupplement

\section{Data and Figures}

\begin{longtable}{|c|c|c|c|c|c|c|}
\hline
$d_\text{Nb}$ (nm) & Pillar length (nm) & Pillar width (nm) & $R_\text{N}^{AP}$ ($\Omega$) & $R_\text{N}^{P}$ ($\Omega$) & $R_\text{S}^{AP}$ ($\Omega$) & $R_\text{S}^{P}$ ($\Omega$) \\
\hline
\endfirsthead
\multicolumn{4}{c}
{\tablename\ \thetable\ -- \textit{Continued from previous page}} \\
\hline
$d_\text{Nb}$ (nm) & Pillar length (nm) & Pillar width (nm) & $R_\text{N}^{AP}$ ($\Omega$) & $R_\text{N}^{P}$ ($\Omega$) & $R_\text{S}^{AP}$ ($\Omega$) & $R_\text{S}^{P}$ ($\Omega$) \\
\hline
\endhead
\hline \multicolumn{4}{r} {\textit{Continued on next page}} \\
\endfoot
\hline
\endlastfoot
0 & 665 & 547 & 0.3143 & 0.3110 &  &  \\ 
0 & 842 & 865 & 0.2075 & 0.2060 &  &  \\ 
0 & 815 & 900 & 0.2285 & 0.2265 &  &  \\ 
24 & 447 & 509 & 0.2378 & 0.2372 &  &  \\ 
24 & 551 & 455 & 0.4469 & 0.4459 &  &  \\ 
24 & 703 & 489 & 0.4374 & 0.4362 &  &  \\ 
24 & 848 & 491 & 0.4004 & 0.3995 &  &  \\ 
24 & 946 & 464 & 0.5198 & 0.5184 &  &  \\ 
14 & 825 & 525 & 0.6227 & 0.6221 &  &  \\ 
14 & 841 & 606 & 0.4799 & 0.4792 &  &  \\ 
14 & 1020 & 651 & 0.5218 & 0.5214 &  &  \\ 
33 & 811 & 388 & 0.6146 & 0.6143 &  &  \\ 
33 & 673 & 370 & 0.7469 & 0.7462 &  &  \\ 
33 & 1280 & 784 & 0.6708 & 0.6707 &  &  \\ 
33 & 1020 & 737 & 0.4965 & 0.4963 &  &  \\ 
33 & 910 & 658 & 0.5152 & 0.5150 &  &  \\ 
19 & 437 & 596 & 0.8595 & 0.8589 &  &  \\ 
14 & 886 & 611 & 0.6072 & 0.6063 &  &  \\ 
14 & 1088 & 551 & 0.6381 & 0.6374 &  &  \\ 
24 & 1111 & 478 & 0.5056 & 0.5051 &  &  \\ 
24 & 1322 & 581 & 0.3868 & 0.3864 &  &  \\ 
24 & 892 & 423 & 0.5230 & 0.5221 &  &  \\ 
24 & 1318 & 617 & 0.3780 & 0.3775 &  &  \\ 
22 & 722 & 541 & 0.5423 & 0.5418 &  &  \\ 
22 & 725 & 539 & 0.6677 & 0.6673 &  &  \\ 
22 & 720 & 527 & 0.5465 & 0.5461 &  &  \\ 
22 & 749 & 693 & 0.5936 & 0.5924 &  &  \\ 
22 & 736 & 615 & 0.7919 & 0.7910 &  &  \\ 
22 & 712 & 479 & 0.7673 & 0.7660 &  &  \\ 
22 & 679 & 416 & 1.1350 & 1.1335 &  &  \\ 
25 & 350 & 468 & 0.7364 & 0.7334 &  &  \\ 
25 & 522 & 457 & 0.7112 & 0.7090 &  &  \\ 
25 & 883 & 474 & 0.6783 & 0.6768 & 0.3242 & 0.3279 \\ 
25 & 1078 & 479 & 0.5995 & 0.5984 & 0.3030 & 0.3055 \\ 
25 & 1315 & 462 & 0.6682 & 0.6670 &  &  \\ 
25 & 1453 & 468 & 0.5252 & 0.5245 &  &  \\ 
19 & 805 & 492 & 0.5450 & 0.5429 &  &  \\ 
19 & 798 & 515 & 0.5140 & 0.5120 &  &  \\ 
19 & 791 & 500 & 0.5955 & 0.5935 &  &  \\ 
19 & 793 & 498 & 0.5319 & 0.5299 &  &  \\ 
19 & 791 & 494 & 0.5364 & 0.5353 &  &  \\ 
19 & 797 & 494 & 0.5378 & 0.5358 &  &  \\ 
21 & 927 & 501 & 0.4155 & 0.4147 &  &  \\ 
21 & 933 & 489 & 0.4376 & 0.4369 &  &  \\ 
21 & 940 & 529 & 0.3983 & 0.3975 &  &  \\ 
21 & 962 & 532 & 0.2836 & 0.2830 &  &  \\ 
21 & 937 & 538 & 0.4597 & 0.4588 &  &  \\ 
29 & 775 & 494 & 0.4893 & 0.4873 & 0.3083 & 0.3069 \\ 
29 & 795 & 482 & 0.4931 & 0.4910 &  &  \\ 
29 & 779 & 499 & 0.5068 & 0.5048 &  &  \\ 
29 & 795 & 525 & 0.4930 & 0.4911 & 0.2995 & 0.2981 \\ 
29 & 750 & 503 & 0.4978 & 0.4958 & 0.3472 & 0.3458 \\ 
29 & 826 & 497 & 0.4676 & 0.4657 & 0.3149 & 0.3134 \\ 
31 & 801 & 501 & 0.4476 & 0.4460 &  &  \\ 
31 & 821 & 521 & 0.4172 & 0.4155 &  &  \\ 
31 & 815 & 494 & 0.4592 & 0.4574 &  &  \\ 
31 & 792 & 501 & 0.5000 & 0.4982 & 0.2809 & 0.3066 \\ 
31 & 799 & 490 & 0.4723 & 0.4713 &  &  \\ 
31 & 810 & 494 & 0.4211 & 0.4193 &  &  \\ 
31 & 812 & 525 & 0.4067 & 0.4050 & 0.1984 & 0.2078 \\ 
21 & 749 & 623 & 0.4538 & 0.4528 &  &  \\ 
21 & 744 & 601 & 0.4754 & 0.4743 &  &  \\ 
21 & 736 & 595 & 0.4950 & 0.4938 &  &  \\ 
21 & 743 & 608 & 0.5490 & 0.5478 &  &  \\ 
21 & 710 & 564 & 0.5692 & 0.5680 &  &  \\ 
21 & 716 & 561 & 0.5308 & 0.5295 &  &  \\ 
21 & 714 & 590 & 0.5374 & 0.5360 & 0.3595 & 0.3797 \\ 
57 & 721 & 639 & 0.4176 & 0.4173 &  &  \\ 
57 & 701 & 618 & 0.3803 & 0.3800 & 0.2560 & 0.2560 \\ 
26 & 705 & 487 & 0.7084 & 0.7076 &  &  \\ 
26 & 696 & 488 & 0.7254 & 0.7246 &  &  \\ 
26 & 717 & 477 & 0.6197 & 0.6190 & 0.4819 & 0.4814 \\ 
26 & 628 & 481 & 1.2160 & 1.2157 &  &  \\ 
26 & 699 & 467 & 0.7153 & 0.7148 &  &  \\ 
26 & 720 & 474 & 0.6588 & 0.6581 & 0.4746 & 0.4738 \\ 
22 & 859 & 443 & 0.5679 & 0.5664 &  &  \\ 
22 & 828 & 463 & 0.5043 & 0.5028 &  &  \\ 
22 & 863 & 424 & 0.4914 & 0.4901 &  &  \\ 
22 & 817 & 470 & 0.4345 & 0.4330 &  &  \\ 
22 & 809 & 475 & 0.3262 & 0.3252 &  &  \\ 
63 & 697 & 495 & 0.3884 & 0.3881 &  &  \\ 
63 & 669 & 468 & 0.5185 & 0.5184 &  &  \\ 
63 & 670 & 505 & 0.4041 & 0.4038 &  &  \\ 
33 & 684 & 493 & 0.5600 & 0.5590 &  &  \\ 
33 & 716 & 504 & 0.5506 & 0.5497 & 0.3744 & 0.3739 \\ 
33 & 705 & 516 & 0.5714 & 0.5705 & 0.4028 & 0.4022 \\ 
33 & 676 & 481 & 0.6019 & 0.6008 &  &  \\ 
33 & 703 & 514 & 0.5554 & 0.5544 &  &  \\ 
33 & 689 & 487 & 0.5927 & 0.5917 &  &  \\ 
33 & 660 & 508 & 0.5345 & 0.5336 & 0.3492 & 0.3486 \\ 
37 & 955 & 350 & 0.4722 & 0.4717 & 0.3784 & 0.3782 \\ 
37 & 977 & 409 & 0.3531 & 0.3526 & 0.2817 & 0.2815 \\ 
37 & 352 & 340 & 0.8040 & 0.8029 &  &  \\ 
41 & 915 & 314 & 0.4918 & 0.4912 &  &  \\ 
41 & 904 & 313 & 0.5724 & 0.5718 & 0.4944 & 0.4941 \\ 
49 & 997 & 523 & 0.3201 & 0.3196 &  &  \\ 
49 & 578 & 541 & 0.2953 & 0.2948 &  &  \\ 
25 & 855 & 432 & 0.5645 & 0.5637 & 0.2570 & 0.2585 \\ 
25 & 726 & 442 & 0.4891 & 0.4877 &  &  \\ 
19 & 970 & 729 & 0.3274 & 0.3268 &  &  \\ 
19 & 971 & 568 & 0.5102 & 0.5095 &  &  \\ 
19 & 957 & 564 & 0.5504 & 0.5498 &  &  \\ 
19 & 973 & 748 & 0.3149 & 0.3144 &  &  \\ 
19 & 962 & 532 & 0.4370 & 0.4364 &  &  \\ 
76 & 354 & 207 & 1.5758 & 1.5755 &  &  \\ 
10 & 497 & 576 & 0.3940 & 0.3931 &  &  \\ 
10 & 436 & 335 & 0.6459 & 0.6451 &  &  \\ 
10 & 528 & 567 & 0.3448 & 0.3442 &  &  \\ 
10 & 510 & 573 & 0.3593 & 0.3582 &  &  \\ 
10 & 640 & 564 & 0.3250 & 0.3245 &  &  \\ 
10 & 418 & 366 & 0.6174 & 0.6152 &  &  \\ 
0 & 452 & 759 & 0.2526 & 0.2480 &  &  \\ 
0 & 512 & 462 & 0.4457 & 0.4398 &  &  \\ 
\hline
\caption{Table of raw data for the devices that appear in the graphs in the main paper and this Supplemental Material. Note that these resistance values include the contribution of the `contact leads' next to the devices. All resistance values are at zero applied field. There are far fewer results in the superconducting state than normal state because many devices, especially for lower $d_\text{Nb}$ values, did not go fully superconducting.}
\label{tab:data}
\end{longtable}

\subsection{Further plots}

\begin{figure}
    \centering
    \includegraphics[width = 0.9\linewidth]{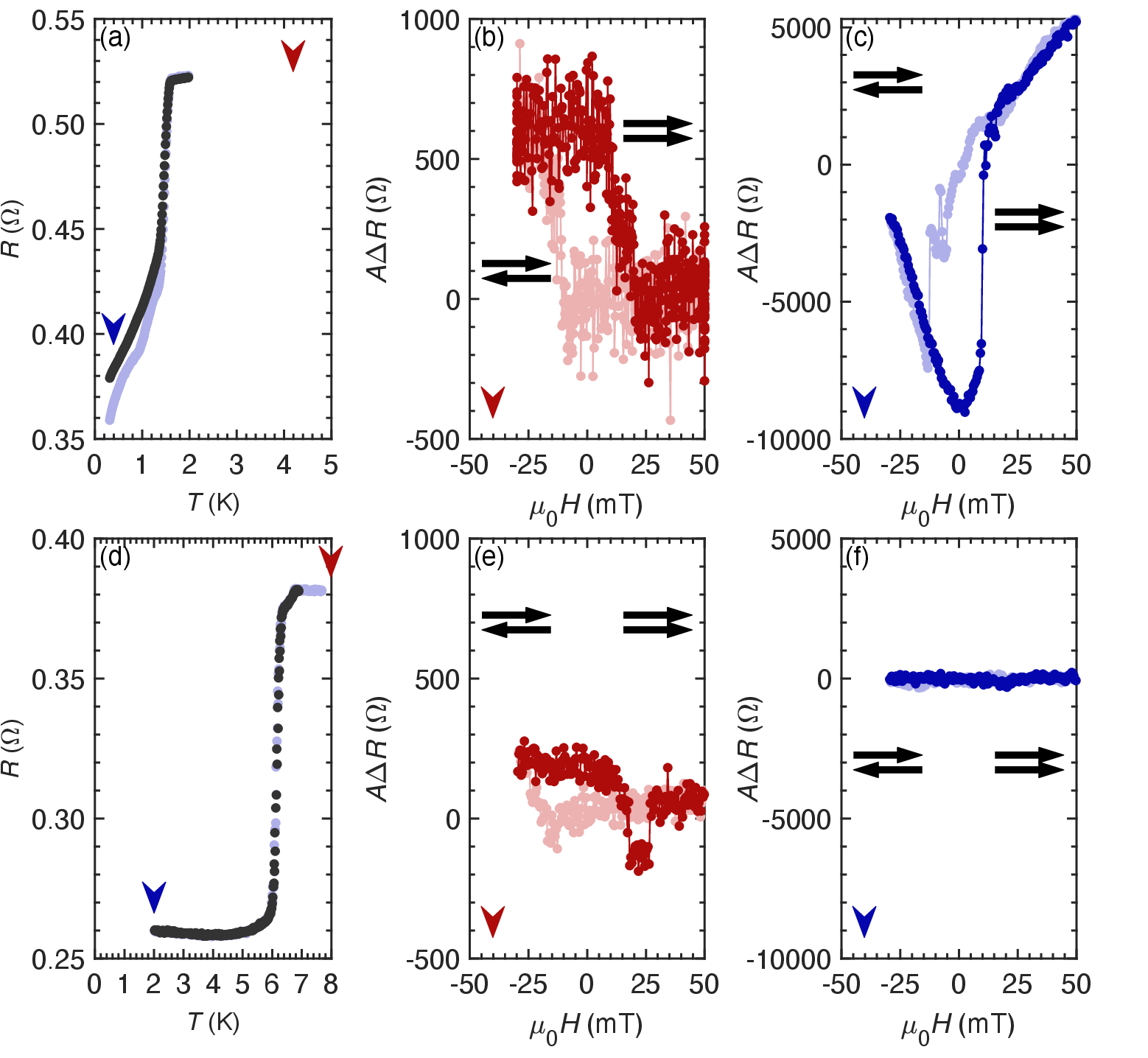}
    \caption{(a,d) $R(T)$ curves for two different devices  (top, $d_\text{Nb} = 21$~nm; bottom, $d_\text{Nb} = 57$~nm) in both the parallel (black) and antiparallel (blue) states. (b,e) Normal state minor $R(H)$ loops from the same devices, both demonstrating GMR. (c,f) Minor $R(H)$ loops in the superconducting state. For the $R(H)$ loops, light data represent sweeps from positive to negative $H$, starting at high positive values.  The dark data in each loop is for the return sweep. In these minor loops, only the free Py layer undergoes switching, illustrated by the black arrows, which show the relative magnetic moment orientation of the Py layers. Arrows in (a,d) indicate the temperature of the corresponding $R(H)$ loops.}
    \label{fig:RHRT}
\end{figure}

Figure~\ref{fig:RHRT} is as Fig.~2 in the main paper, showing $R(T)$ and $R(H)$ loops for two more devices, which represent extremes of the two different behaviours observed in these papers. The top graphs come from the device with $d_\text{Nb} = 21$~nm, which is the thinnest $d_\text{Nb}$ tested. This device demonstrates superconducting spin valve effect behaviour even in the wires, visible in Fig.~\ref{fig:RHRT}(a), as the $T_\text{c}$ of the parallel state is lower than that of the antiparallel state in the steep wires region. The device transition is noticeably extremely wide. The bottom graphs come from a device with $d_\text{Nb} = 57$~nm, which is large, as reflected in the small values of $A\Delta R$ in the $R(H)$ loops and the negligible $\Delta T_\text{c}$ in Fig.~\ref{fig:RHRT}(d).\par
\begin{figure}
    \centering
    \includegraphics[width = 0.9\linewidth]{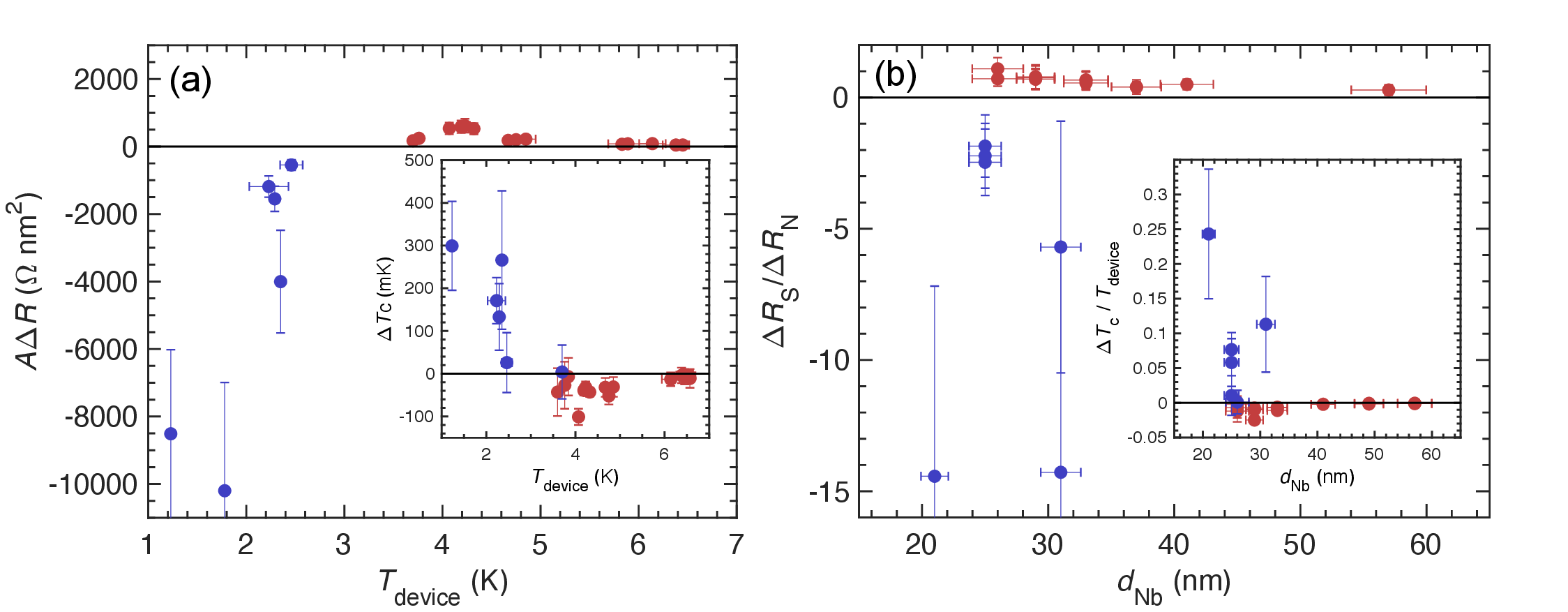}
    \caption{(a) $A\Delta R$ ($T/T_\text{device}= 0.3$ ) and inset $\Delta T_\text{c}$ vs. $T_\text{device}$. The crossover is still present, and points that were outliers when plotted vs. $d_\text{Nb}$ now fit the trend better. (b) $\Delta R$ at $T/T_\text{device}= 0.3$ normalised by $\Delta R$ in the normal state. Inset: $\Delta T_\text{c}$ normalised by $T_\text{device}$.}
    \label{fig:crossover}
\end{figure}

Figure~\ref{fig:crossover}(a) shows $A\Delta R$ at $T/T_\text{device}= 0.3$ and $\Delta T_\text{c}$ as in the main text, but plotted against $T_\text{device}$. The outlier points at -11,000 and -4,000 $\Omega$\,nm\textsuperscript{2} agree with the main trend much better when this is used as the independent parameter, which suggests that the properties of the superconductor, not just the spacing between the ferromagnets, influence the crossover behaviour. Figure~\ref{fig:crossover}(b) shows normalised data, $\Delta R_\text{S} / \Delta R_\text{N}$ where $\Delta R_\text{S}$ is $\Delta R$ at $T/T_\text{device}= 0.3$, and $\Delta R_\text{N}$ is $\Delta R$ in the normal state, along with $\Delta T_\text{c} / T_\text{device}$ in the inset. The normalised data also show a sign change dependent on $d_\text{Nb}$, and the crossover occurs at a similar value of $d_\text{Nb}$ to the standard data. This indicates the scatter shown by the devices is not responsible for the crossover and trend we see below $T_\text{c}$.

\section{Data analysis}
\subsection{Link between $\Delta R$ and $\Delta T_\text{c}$}
\begin{figure}
    \centering
    \includegraphics[width = 0.5\linewidth]{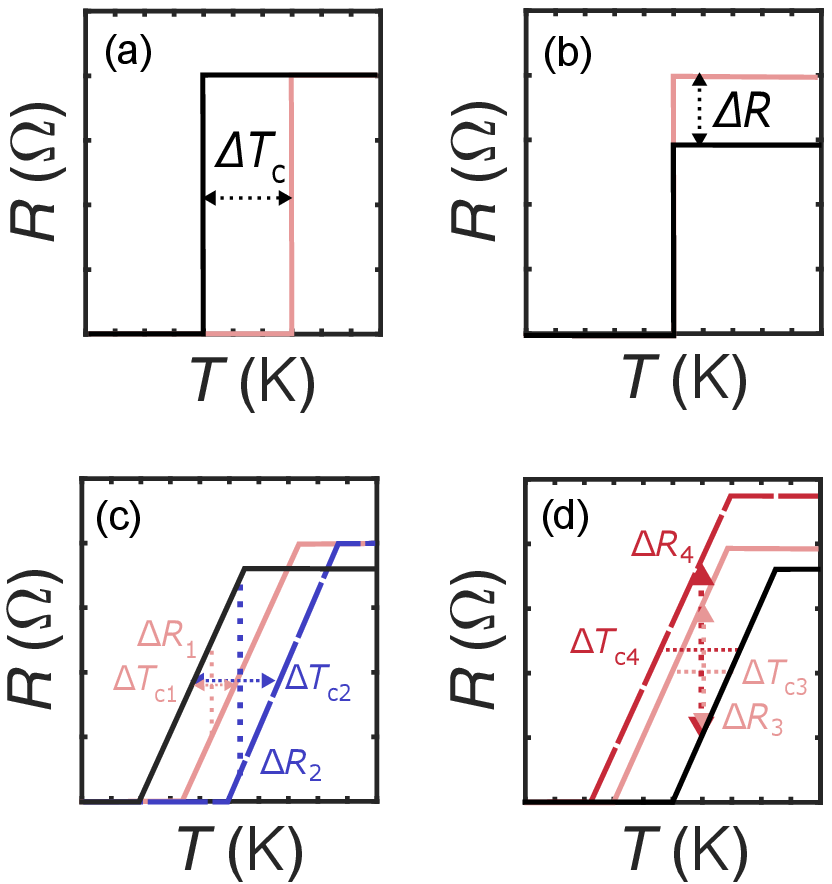}
    \caption{Figures showing how the superconducting transition leads to a link between $\Delta R$ and $\Delta T_\text{c}$ values, exaggerated for clarity. (a) In an ideal device showing only the superconductor proximity effect, the parallel state (black) has lower $T_\text{c}$ than the antiparallel (pink) state ($\Delta T_\text{c} > 0$). (b) In an ideal device showing only GMR, the antiparallel state has a higher resistance above $T_\text{c}$ than the parallel state ($\Delta R > 0$). (c) A finite gradient in the superconducting transition links $\Delta T_\text{c}$ and $\Delta R$. Here, for the same normal state $\Delta R$, $\Delta T_\text{c2} > \Delta T_\text{c1}$ ($\Delta T_\text{c2}$ is more positive) causing $\Delta R_2$ in the transition to be more negative than $\Delta R_1$. (d) Inversely, a larger $\Delta R$ in the normal state can lead to a more negative $\Delta T_\text{c}$: here $\Delta R_4 > \Delta R_3 $, so $\Delta T_\text{c4}$ appears to have a greater (more negative) value than $\Delta T_\text{c3}$.}
    \label{fig:RTlink}
\end{figure}

Real (non-ideal) devices have a superconducting transition with a finite gradient, which leads to a link between  $\Delta R$ and $\Delta T_\text{c}$ within this region, where the measurement of $\Delta T_\text{c}$ occurs. Hence, as shown in Fig.~\ref{fig:RTlink}, effects causing a shift in either $\Delta R$ or $\Delta T_\text{c}$ can cause a shift in the other parameter as well. This must be accounted for when considering devices with both the superconducting spin valve effect, which affects $\Delta T_\text{c}$, and GMR, which affects $\Delta R$. 

\subsection{$\Delta T_\text{c}$ extraction}
$\Delta T_\text{c}$ values were extracted from separate $R(T)$ measurements of the parallel and antiparallel states of the devices, at $R(T_\text{c}) = 0.5R_\text{N}^{AP}$, where $R_\text{N}^{AP}$ is the resistance at $T_\text{device}$, the onset of the device transition in the antiparallel state. A single $0.5R_\text{N}$ value is used for both states as this gives a more representative value of the temperature instability allowed for a working device. \par
\begin{figure}
    \centering
    \includegraphics[width = 0.9\linewidth]{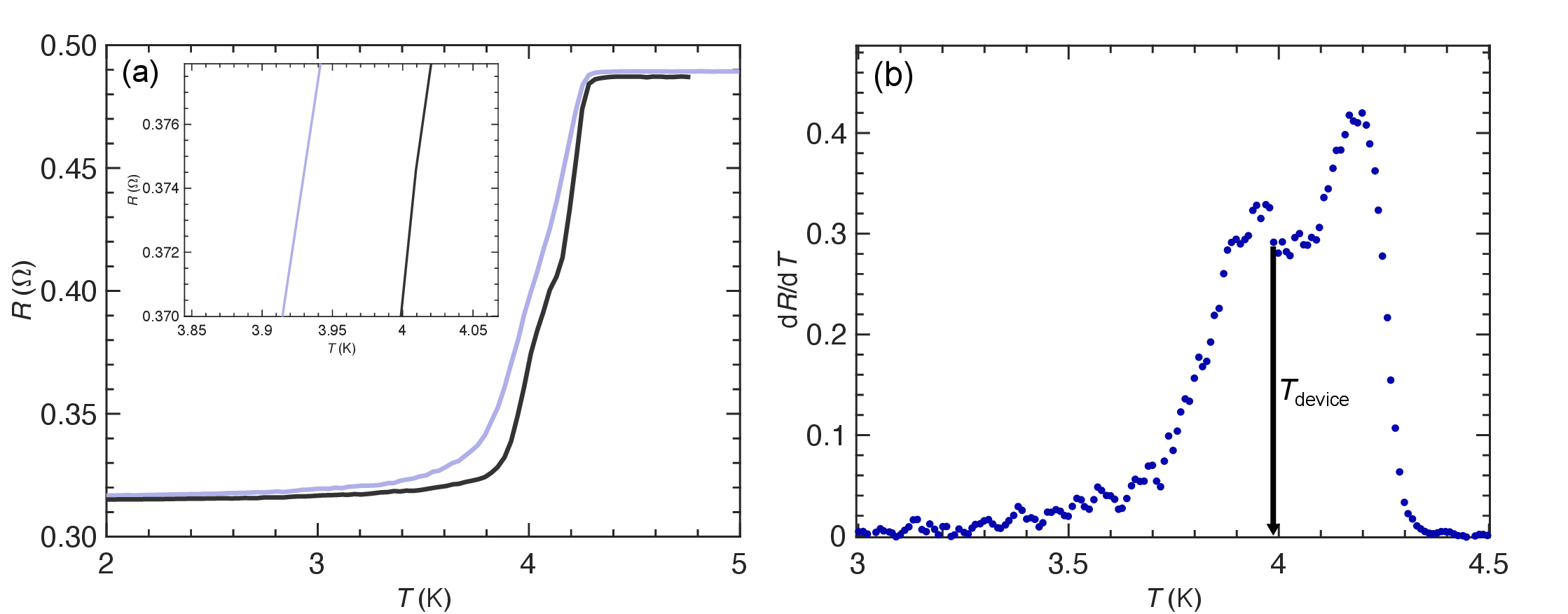}
    \caption{(a) $\Delta T_\text{c}$ is extracted from raw $R(T)$ data in the parallel (black) and antiparallel (light blue) curves. The curves are zoomed in near $R_\text{N}^{AP}$ (inset) and $\Delta T_\text{c}$ is the difference between the temperatures corresponding to this resistance for each curve. (b) $T_\text{device}$ is taken as the temperature at which the gradient of the antiparallel curve increases by 0.03 due to the device transition.}
    \label{fig:Tc_extraction}
\end{figure}

 
Temperature sweeps were performed at a rate of 0.3~K/min, using PID control. Higher cooling rates gave the same value of $T_\text{c}$, but this lower rate was used to ensure accuracy. $R(T)$ sweeps from high to low temperatures are binned and averaged to minimise the effect of random noise in $R$. The averaged parallel and antiparallel state curves are plotted together and the $T_\text{c}$ value for each curve taken at $0.5R_\text{N}$, so that $\Delta T_\text{c}=T_\text{c}^{AP}-T_\text{c}^{P}$ [Fig.~\ref{fig:Tc_extraction}(a)]. Temperature values are also taken at $0.45R_\text{N}$ and $0.55R_\text{N}$ which provide errors in each $T_\text{c}$ measurement. The errors of each $T_\text{c}$ are added in quadrature to give the error in $\Delta T_\text{c}$. $T_\text{device}$ is the onset of the superconducting transition for the device. This is calculated from the gradient of the antiparallel state $R(T)$ curve, which is binned and smoothed by a moving average. \par $T_\text{device}$ is the temperature at which the gradient change associated with the device transition first exceeds 0.03 [Fig.~\ref{fig:Tc_extraction}(b)]. This value of 0.03 was chosen as a value exceeding the basic noise level of the gradient data. Gradient is calculated from binned data and smoothed by a moving median. Errors in $T_\text{device}$ are the range of possible values that satisfy this criterion.

\bibliography{bib.bib}